\documentclass[aps,longbibliography,eqsecnum, twocolumn]{revtex4-1}
\usepackage{graphicx}
\usepackage{stmaryrd}
\usepackage{amsmath, amsfonts}
\usepackage{mathrsfs}
\usepackage{braket}
\usepackage{color}
\usepackage{xspace}
\usepackage{amscd}
\usepackage{comment}
\usepackage{bm}
\definecolor{lightblue}{rgb}{0.13, 0.26, 0.99}

\usepackage[
colorlinks=true,
urlcolor=blue,
citecolor=blue,
linkcolor=blue,
hyperfootnotes=false]{hyperref}

\newcommand{\tr}{\operatorname{tr}}

\newcommand{\bacovo}{$\mathrm{BaCo_2V_2O_8}$}
\newcommand{\srcovo}{$\mathrm{SrCo_2V_2O_8}$}
\newcommand{\cupm}{$\mathrm{[Cu(pym)(H_2O)_4]SiF_6 \cdot H_2O}$}

\allowdisplaybreaks

\begin{document}

\title{Field-induced dimer orders in quantum spin chains}
\author{Shunsuke C. Furuya}
\affiliation{Condensed Matter Theory Laboratory, RIKEN, Wako, Saitama 351-0198, Japan}

\begin{abstract}
    Field-induced excitation gaps in quantum spin chains are an interesting phenomenon related to confinements of topological excitations. 
    In this paper, I present a novel type of this phenomenon.
    I show that an effective magnetic field with a fourfold screw symmetry induces the excitation gap accompanied by dimer orders.
    The dimer order parameter and the excitation gap exhibit characteristic power-law dependence on the fourfold screw-symmetric field.
    Moreover, the field-induced dimer order and the field-induced  N\'eel order coexist when the external uniform magnetic field, the fourfold screw-symmetric field, and the twofold staggered field are applied.
    This situation is in close connection with a compound \cupm{} [J. Liu \textit{et al.}, Phys. Rev. Lett. {\bf 122}, 057207 (2019)].
    In this paper, I discuss a mechanism of field-induced dimer orders by using a density-matrix renormalization group method, a perturbation theory, and quantum field theories.
\end{abstract}

\date{\today}
\maketitle

\section{Introduction}\label{sec:intro}

Quantum spin-$1/2$ chains do not have a unique gapped ground state in the presence of the time-reversal symmetry unless either the U(1) spin-rotation symmetry or the translation symmetry is broken~\cite{lsm, affleck_nonabelian, furuya_wzw}.
For example, the spin-$1/2$ Heisenberg antiferromagnetic (HAFM) chain has a unique gapless ground state called the Tomonaga-Luttinger (TL) liquid state~\cite{giamarchi_book}.
Even when the time-reversal symmetry is broken by the external magnetic field, the TL liquid does not immediately acquire the gap, though it eventually does with the saturated magnetization~\cite{metlitski_lsm}.
This is because the external magnetic field is uniform in the scale of spin chains.
It breaks neither the U(1) rotation nor the translation symmetry.
Interestingly, however, there are several spin-$1/2$ chain compounds where the magnetic field immediately opens the excitation gap~\cite{dender_cubenzoate, zvyagin_cupm_2004, umegaki_kcugaf6}.

This puzzle of the field-induced gap was found in Cu Benzoate~\cite{dender_cubenzoate} and later solved with quantum field theories~\cite{oshikawa_stag_prl, affleck_fig, furuya_bbs}.
Essentially, the field-induced excitation gap in those compounds comes from an absence of a bond-centered inversion symmetry.
This low crystalline symmetry allows the $g$ tensor of electron spins to have a
twofold staggered component.
The magnetic field, when combined with the low symmetry, generates a twofold staggered magnetic field that breaks the translation symmetry.
As a result, the uniform magnetic field induces the excitation gap and also the N\'eel order in the direction of the effectively generated twofold staggered magnetic field.
Thanks to the dimensionality and strong interactions among elementary excitations, the excitation gap and the N\'eel order exhibit interesting power-law behaviors that deviate from spin-wave predictions~\cite{oshikawa_stag_prl, affleck_fig, furuya_cnlsm}.
The phenomenon of the field-induced excitation gap has drawn attention for its connection with confinement of topological excitations~\cite{faure_bacovo, takayoshi_bacovo}.

In this paper, I discuss a novel field-induced excitation gap phenomenon.
That is field-induced dimer orders in quantum spin chains.

\begin{figure}[t!]
    \centering
    \includegraphics[viewport = 0 0 2000 2000, width=\linewidth]{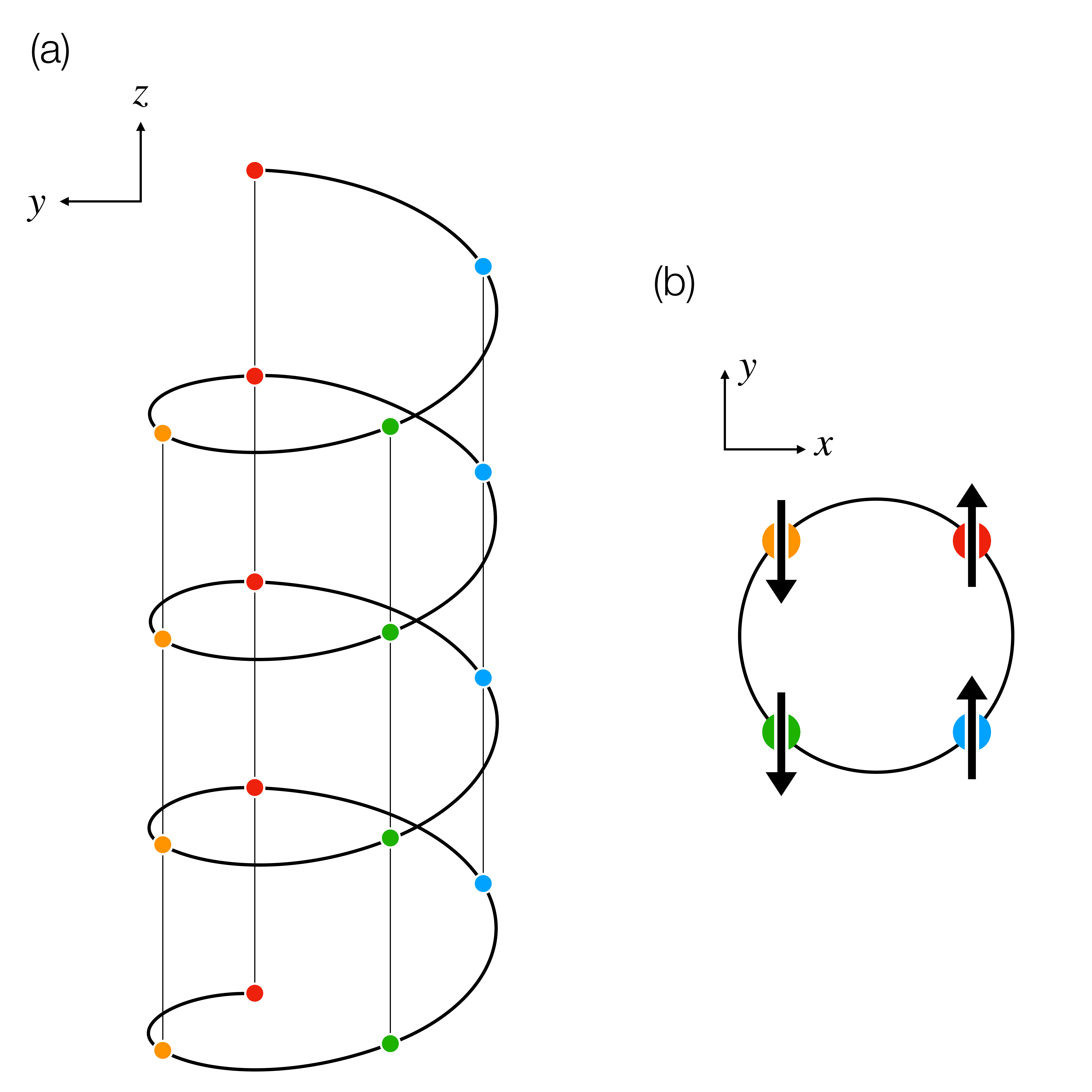}
    \caption{(a) A fourfold screw spin chain. Each ball represents a spin operator. (b) A fourfold screw field symmetric under the bond-centered inversion $I_b$. Arrows depict directions of the fourfold screw field. }
    \label{fig:screw}
\end{figure}

A key ingredient is a fourfold screw symmetry (Fig.~\ref{fig:screw}).
The screw structure can violate two kinds of inversion symmetries at the same time, namely, the bond-centered inversion symmetry and a site-centered inversion symmetry.
Such a fourfold structure is indeed incorporated in the $g$ tensor of spin-chain compounds, \bacovo{}~\cite{kimura_bacovo_jpsj, faure_bacovo} and \cupm{}~\cite{liu_chain_esr}.
When the uniform magnetic field is externally applied, the fourfold screw symmetry manifests itself as an effective fourfold screw-symmetric magnetic field.
The fourfold screw field brings dimer orders to spin chains immediately.
I discuss first a mechanism of the dimer-order generation.
Next I take the twofold staggered field into account and discuss coexistent growth of the dimer and N\'eel orders with increase of the uniform magnetic field.

This paper is organized as follows.
I define a spin-chain model and show numerical evidence of the field-induced dimer orders in the simplest case in Sec.~\ref{sec:screw}.
A qualitative mechanism of field-induced dimer orders is discussed in Sec.~\ref{sec:xy}, where the spin-chain model is replaced to a spinless fermion model which is smoothly deformed from the original spin-chain model.
Here, the low-energy effective Hamiltonian is systematically derived.
In Sec.~\ref{sec:bosonization}, on the basis of observations made in Sec.~\ref{sec:xy}, I develop a quantum field theory that explains quantitatively numerical results of Sec.~\ref{sec:screw}.
The quantum field theory also predicts the coexistence of the dimer and N\'eel orders both of which grow with the uniform magnetic field.
This coexistent growth of the dimer and N\'eel orders are discussed in Sec.~\ref{sec:coex}, which is supported by numerical calculations.
I also discuss relevance of theoretical results to experiments in Sec.~\ref{sec:cupm}.
Finally, I summarize the paper in Sec.~\ref{sec:summary}.

\begin{figure}[t!]
    \centering
    \includegraphics[viewport = 0 0 864 504, width=\linewidth]{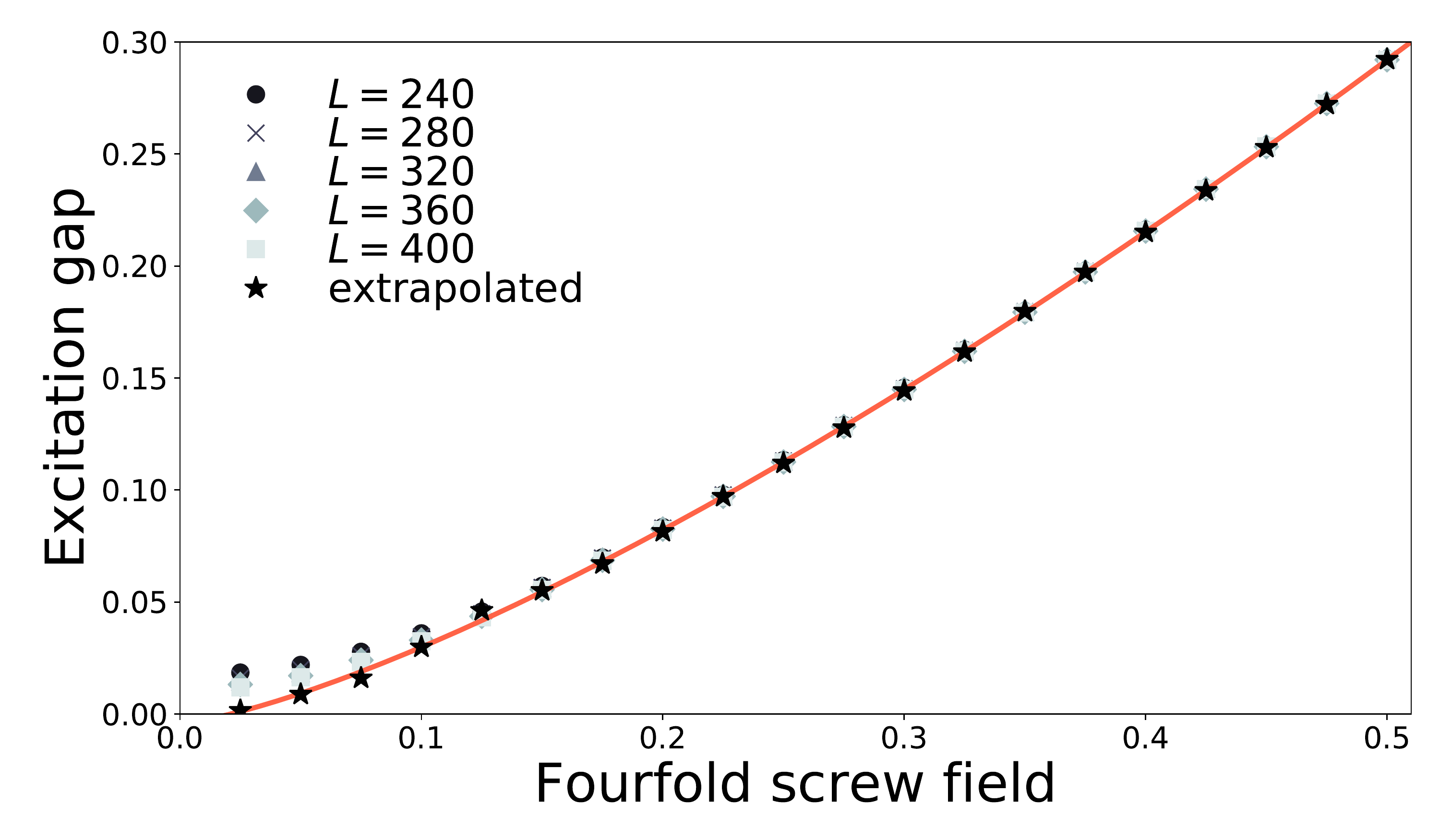}
    \caption{The lowest-energy excitation gap from the ground state of the Hamiltonian \eqref{H4} is plotted against the fourfold screw field $h_4$ for system sizes $L$ ranging from $L=240$ to $400$.
    The gap is extrapolated to the $L\to +\infty$ limit by using a formula $\Delta = a_0 + \frac{a_1}L + \frac{a_2}{L^2}$, where $a_n$ ($n=0,1,2$) are fitting parameters.
    The error of the extrapolated data is estimated to be  $11~\%$ for $h_4/J=0.05$ and $<0.28~\%$ for $h_4/J \ge 0.1$.
    The solid curve is the best fit of the extrapolated data by a function $\Delta = b_0{h_4}^{b_1} +b_2$ with fitting parameters $b_n$ ($n=0,1,2$).
    Though the fitted result has an unphysical offset $b_2\not=0$, its $h_4$ dependence implies the gap $\Delta \propto J(h_4/J)^{1.34}$. 
    This estimation of the agrees with the field-theoretical prediction \eqref{gap_4/3}.
    }
    \label{fig:gap-vs-hb}
\end{figure}

\section{Screw field}\label{sec:screw}

\subsection{Definition of the model}

In this paper I discuss a quantum spin-$1/2$ chain with the following Hamiltonian:
\begin{align}
    \mathcal H
    &= J \sum_j \bm S_j \cdot \bm S_{j+1} - h_0 \sum_j S_j^z - h_2 \sum_j (-1)^j S_j^x
    \notag \\
    &\qquad - h_4 \sum_j \delta_j S_j^z,
    \label{H_def}
\end{align}
where $\bm S_j$ is the $S=1/2$ spin operator, $J>0$ is the antiferromagnetic exchange coupling, and $\delta_j =1, 1, -1, -1$ for respectively, $j=0,1,2,3 \mod 4$.
Parameters $h_0$, $h_2$, and $h_4$ denote the uniform magnetic field, the twofold staggered field, and the fourfold screw field, respectively.
Note that  $\delta_j$ has a simple expression,
\begin{align}
    \delta_j &= \sqrt{2}\cos\biggl(\pi \frac{2j-1}{4}\biggr).
    \label{delta_j}
\end{align}
Throughout this paper, I employ the unit of $\hbar = a = 1$ unless otherwise stated, where $a$ is the lattice spacing.

The spin-chain model \eqref{H_def} is related to a model proposed for \cupm{}~\cite{liu_chain_esr} but differs from the latter in three points: field directions, a weak exchange anisotropy, and a uniform Dzyaloshinskii-Moriya (DM) interaction.
In Ref.~\cite{liu_chain_esr}, $h_0$ and $h_2$ are applied in the $x$ and the $y$ directions, respectively.
The model used in Ref.~\cite{liu_chain_esr} contains a weak XXZ interaction and the uniform DM interaction.
Those differences hardly affect the ground state of \cupm{}, which are to be clarified in Sec.~\ref{sec:cupm}.

In the compound \cupm{}~\cite{liu_chain_esr}, the twofold staggered field $h_2$ and the fourfold screw field $h_4$ originate from the $g$ tensor of electrons and are thus proportional to the externally applied uniform magnetic field $h_0$.
In this section, I first deal with an unrealistic but simplest situation with $h_0=h_2=0$ and $h_4\not=0$ in Sec.~\ref{sec:h4}.
I will discuss a more realistic situation with $h_2\propto h_0$ and $h_4\propto h_0$ later in Sec.~\ref{sec:coex}.

\subsection{Fourfold screw field}\label{sec:h4}

\begin{figure}[t!]
    \centering
    \includegraphics[viewport = 0 0 865 504, width=\linewidth]{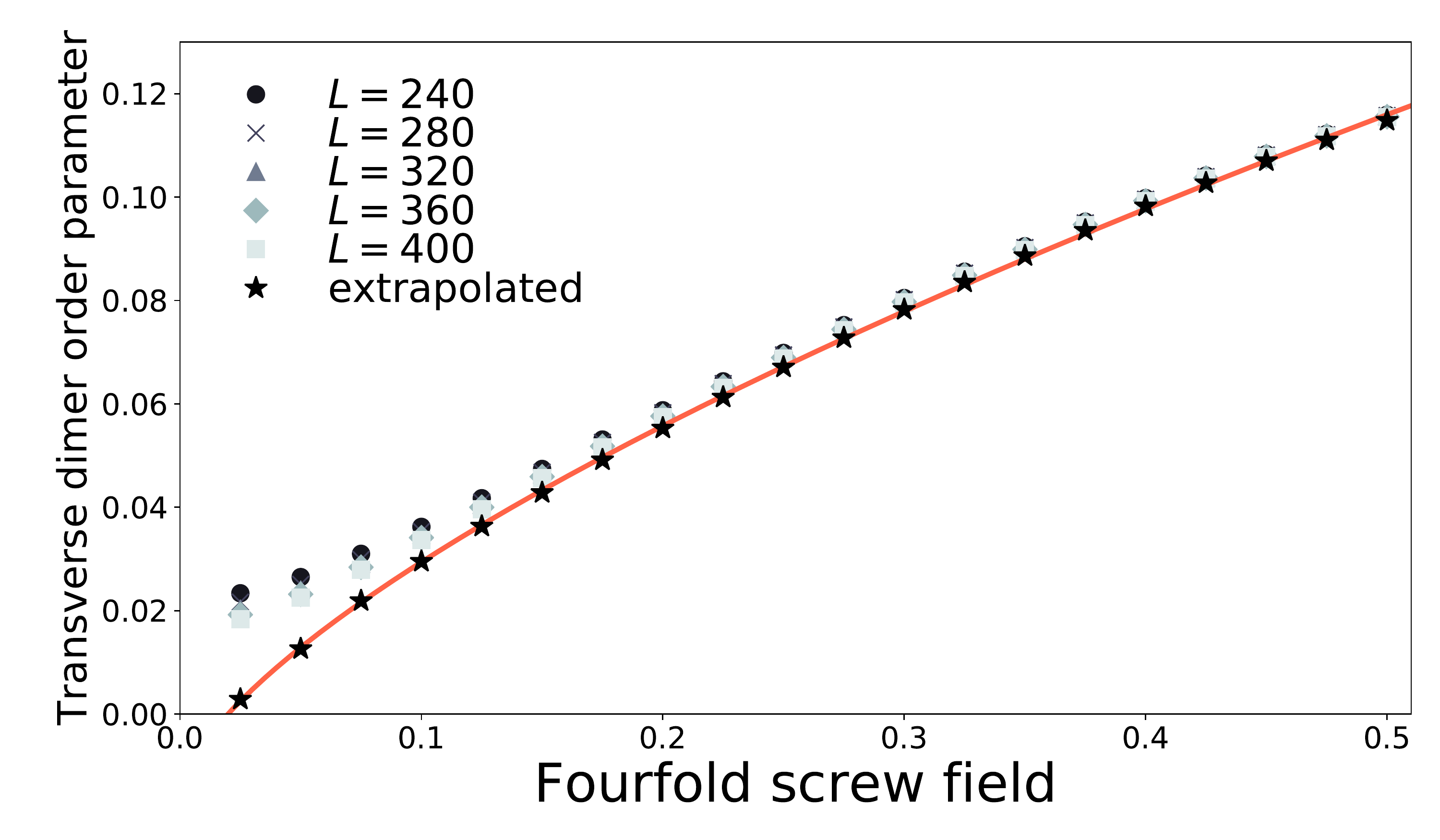}
    \caption{The transverse dimer order parameter \eqref{D_perp} of the ground state of the Hamiltonian \eqref{H4} is plotted against the fourfold screw field $h_4$ for system sizes from $L=240$ to $400$. 
    The dimer is extrapolated to the $L\to +\infty$ limit by using a formula $D_\perp =a'_0 + \frac{a'_1}{\sqrt L} + \frac{a'_2}{L}$ because the scaling dimension of the dimer order parameter is $1/2$.
    The error of the extrapolated data is estimated to be  $6.5~\%$ for $h_4/J=0.025$ and $<0.21~\%$ for $h_4/J\ge 0.1$.
    The error monotonically decreases with incraese of $h_4/J$.
    The solid curve is the best fit of the extrapolated data by a function $D_\perp = b'_0{h_4}^{b'_1} +b'_2$ with fitting parameters $b'_n$ ($n=0,1,2$).
    Though $D_\perp$ shows an unphysical offset $b'_2\not=0$, it implies a power law $D_\perp \propto (h_4/J)^{0.672}$.
    It agrees with the field-theoretical prediction \eqref{xdimer_2/3}.}
    \label{fig:xydimer-vs-hb}
\end{figure}
\begin{figure}[t!]
    \centering
    \includegraphics[viewport = 0 0 865 504, width=\linewidth]{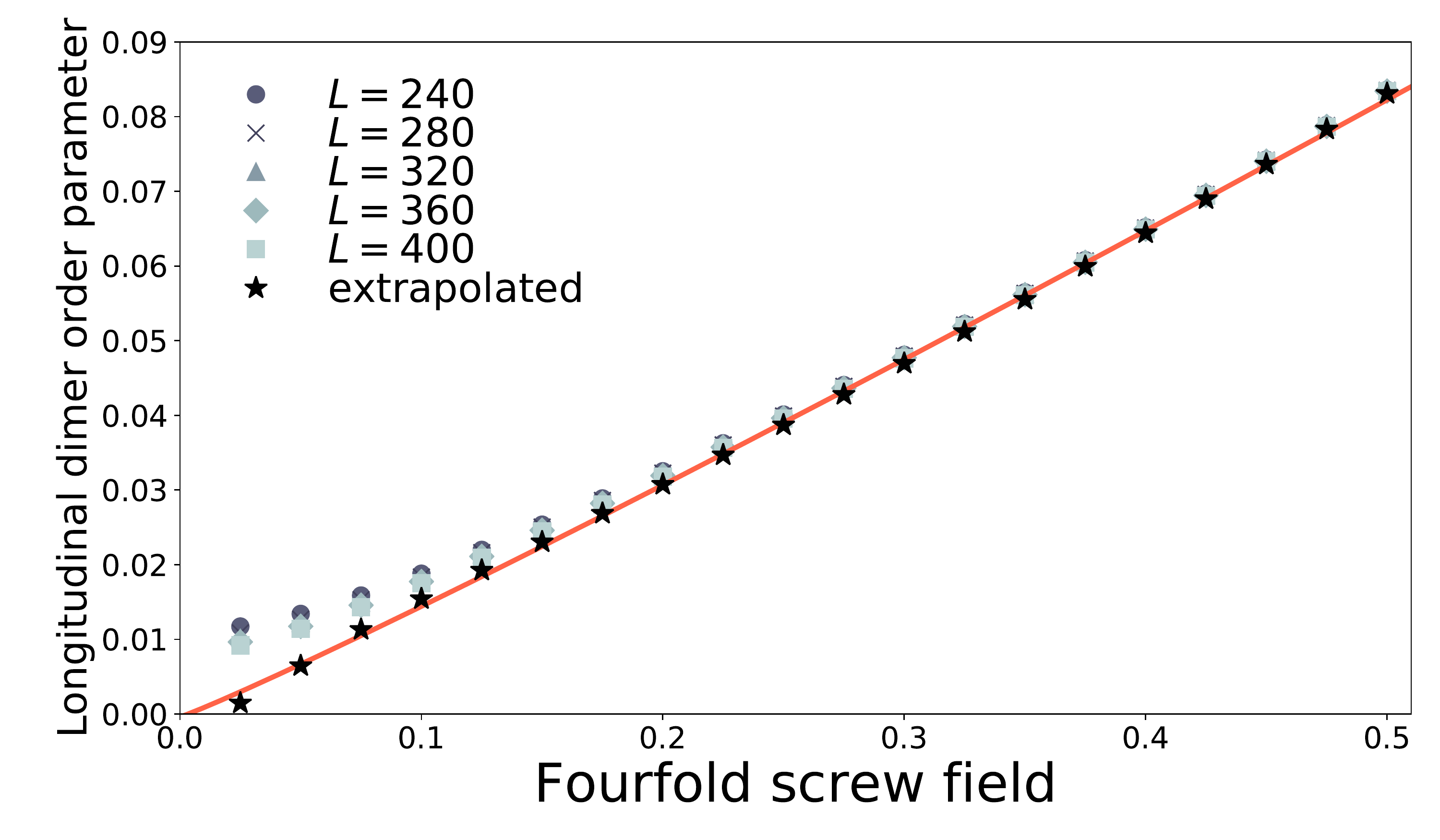}
    \caption{The longitudinal dimer order parameter \eqref{D_parallel} of the ground state of the Hamiltonian \eqref{H4} is plotted against the fourfold screw field $h_4$ for system sizes from $L=240$ to $400$. 
    The dimer is extrapolated to the $L\to +\infty$ limit in the same way as $D_\perp$.
    The error of the extrapolated data is estimated to be  $6.4~\%$ for $h_4/J=0.025$ and $<0.20~\%$ for $h_4/J\ge 0.1$.
    The solid curve is the best fit of the extrapolated data.
    It implies a power law $D_\parallel \propto (h_4/J)^{1.07}$, which differs from that for $D_\perp$ [Eq.~\eqref{D_perp_num}].}
    \label{fig:zdimer-vs-hb}
\end{figure}

The fourfold screw field $h_4$ can generate an excitation gap all by itself to the ground state of the spin chain.
To show this, I set $h_0 = h_2 = 0$ and discuss the $h_4$ dependence of the lowest-energy excitation gap from the ground state.
The Hamiltonian is thus simplified as
\begin{align}
    \mathcal H_4 &= J \sum_j \bm S_j \cdot \bm S_{j+1} -h_4 \sum_j \delta_j S_j^z.
    \label{H4}
\end{align}
When $h_4=0$, the ground state of the model \eqref{H4} is gapless~\cite{giamarchi_book}.
Figure~\ref{fig:gap-vs-hb} shows numerical results on the excitation gap obtained by using the density-matrix renormalization group (DMRG) method with the ITensor C++ library~\cite{ITensor}, where I used the bond dimension $\chi = 400$ and the truncation error cutoff $1\times 10^{-10}$.
Note that all the DMRG calculations in this paper were performed with the open boundary condition.
The DMRG result implies that an infinitesimal $h_4/J$ immediately opens the excitation gap between the ground state and the lowest-energy excited state,
\begin{align}
    \Delta \propto J \biggl(\frac{h_4}{J}\biggr)^{1.34}.
    \label{gap_num}
\end{align}
Similarly to the twofold staggered field, the fourfold screw field induces an excitation gap with a power law.
However, the power $1.34$ differs from that, $2/3$, of the twofold staggered field~\cite{oshikawa_stag_prl, affleck_fig}.

Unlike the twofold staggered field, the fourfold screw field $h_4$ induces no N\'eel order.
Instead, $h_4$ induces dimer orders (Figs.~\ref{fig:xydimer-vs-hb} and \ref{fig:zdimer-vs-hb}),
\begin{align}
    D_\perp 
    &= \frac 1L \sum_j \braket{(-1)^j (S_j^x S_{j+1}^x + S_j^y S_{j+1}^y)},
    \label{D_perp} \\
    D_\parallel
    &= \frac 1L \sum_j \braket{(-1)^j S_j^z S_{j+1}^z},
    \label{D_parallel}
\end{align}
where $L$ is the length of the spin chain.
The dimer order parameters \eqref{D_perp} and \eqref{D_parallel} show different power-law dependence on $h_4$.
DMRG results (Figs.~\ref{fig:xydimer-vs-hb} and \ref{fig:zdimer-vs-hb}) imply
\begin{align}
    D_\perp &\propto \biggl(\frac{h_4}{J}\biggr)^{0.672},
    \label{D_perp_num} \\
    D_\parallel &\propto \biggl(\frac{h_4}{J}\biggr)^{1.07}.
    \label{D_parallel_num}
\end{align}

Induction of $D_\parallel$ by $h_4$ is easily understandable.
Let us recall that $h_4$ is coupled to an operator,
\begin{align}
    f_j^z &= \sqrt{2} \cos\biggl(\pi \frac{2j-1}{4}\biggr) S_j^z.
    \label{fj_def}
\end{align}
The fourfold screw field induces the uniform $f^z$ order:
\begin{align}
    \sum_j \braket{f_j^z}\not=0.
    \label{fz}
\end{align}
The longitudinal dimer order parameter $D_\parallel$ is written in terms of $f_j^z$ as
\begin{align}
    D_\parallel &= \frac 1L \sum_j \braket{f_j^z f_{j+1}^z}.
\end{align}
Nonzero $D_\parallel$ follows immediately from the uniform $f_j^z$ order \eqref{fz}.
However, the induction of the transverse dimer order \eqref{D_perp_num} is nontrivial.

\section{Free spinless fermion theory}\label{sec:xy}

This section is devoted to a qualitative explanation on a mechanism of the field-induced transverse dimer order \eqref{D_perp_num}.
For this purpose, I rewrite the spin chain \eqref{H4} in terms of spinless fermions with the aid of the Jordan-Wigner transformation~\cite{giamarchi_book}.
Let $c_j^\dag$ and $c_j$ be creation and annihilation operators of the spinless fermion at the site $j$, respectively.
The spin-chain model \eqref{H4} is equivalent to the following model of interacting spinless fermions:
\begin{align}
    \mathcal H_4
    &= -\frac J2 \sum_j (c_j^\dag c_{j+1} + \mathrm{H.c.})  - h_4 \sum_j \delta_j \biggl( c_j^\dag c_j - \frac 12 \biggr)
    \notag \\
    &\qquad + J \sum_j \biggl(c_j^\dag c_j - \frac 12\biggr)\biggl(c_{j+1}^\dag c_{j+1} - \frac 12\biggr),
    \label{H4_fermion}
\end{align}
where $\mathrm{H.c.}$ denotes the Hermitian conjugate.

The interaction of spinless fermions, the second line of Eq.~\eqref{H4_fermion}, comes from the longitudinal component of the exchange interaction, $J\sum_j S_j^z S_{j+1}^z$.
Even if this interaction is ignored, qualitative aspects of the ground state are kept intact since the HAFM chain and the XY chain belong to the same TL-liquid phase~\cite{giamarchi_book}.
Therefore, I discuss in this section the free spinless fermion model, 
\begin{align}
    \mathcal H_{\rm XY}
    &= - J \sum_j (c_j^\dag c_{j+1} + \mathrm{H.c.}) - h_4\sum_j \delta_j \biggl(c_j^\dag c_j - \frac 12 \biggr),
    \label{H_XY_fermion}
\end{align}
instead of the model \eqref{H4_fermion}.
In terms of spins, it is the XY chain in the fourfold screw field,
\begin{align}
    \mathcal H_{\rm XY} &= J \sum_j (S_j^x S_{j+1}^x + S_j^y S_{j+1}^y ) +h_4 \sum_j \delta_j S_j^z.
    \label{H_XY}
\end{align}

\subsection{Particle-hole excitations}

Performing the Fourier transformation on Eq.~\eqref{H_XY}, I obtain 
\begin{align}
    \mathcal H_{\rm XY}
    &= \sum_k \epsilon(k) c_k^\dag c_k
    \notag \\
    &\qquad -\frac{h_4}{\sqrt 2} \sum_k \bigl(
    e^{-\pi i/4} c_k^\dag c_{k+\frac \pi 2} + e^{\pi i/4} c_{k+\frac \pi 2}^\dag c_k \bigr)
    \notag \\
    &\qquad + \mathrm{const.}
    \label{H_XY_k}
\end{align}
where $\epsilon(k) = -(J/2) \cos k$ and  $k \in (-\pi, \pi]$ is the wave number.
When $J_4=0$, the spinless fermion is free and has the simple cosine dispersion.
Since the total magnetization is zero in the XY chain, the cosine band is half occupied and the Fermi points are located at $\pm \pi /2$.

\begin{figure}[t!]
    \centering
    \includegraphics[viewport = 0 0 1920 1080, width=\linewidth]{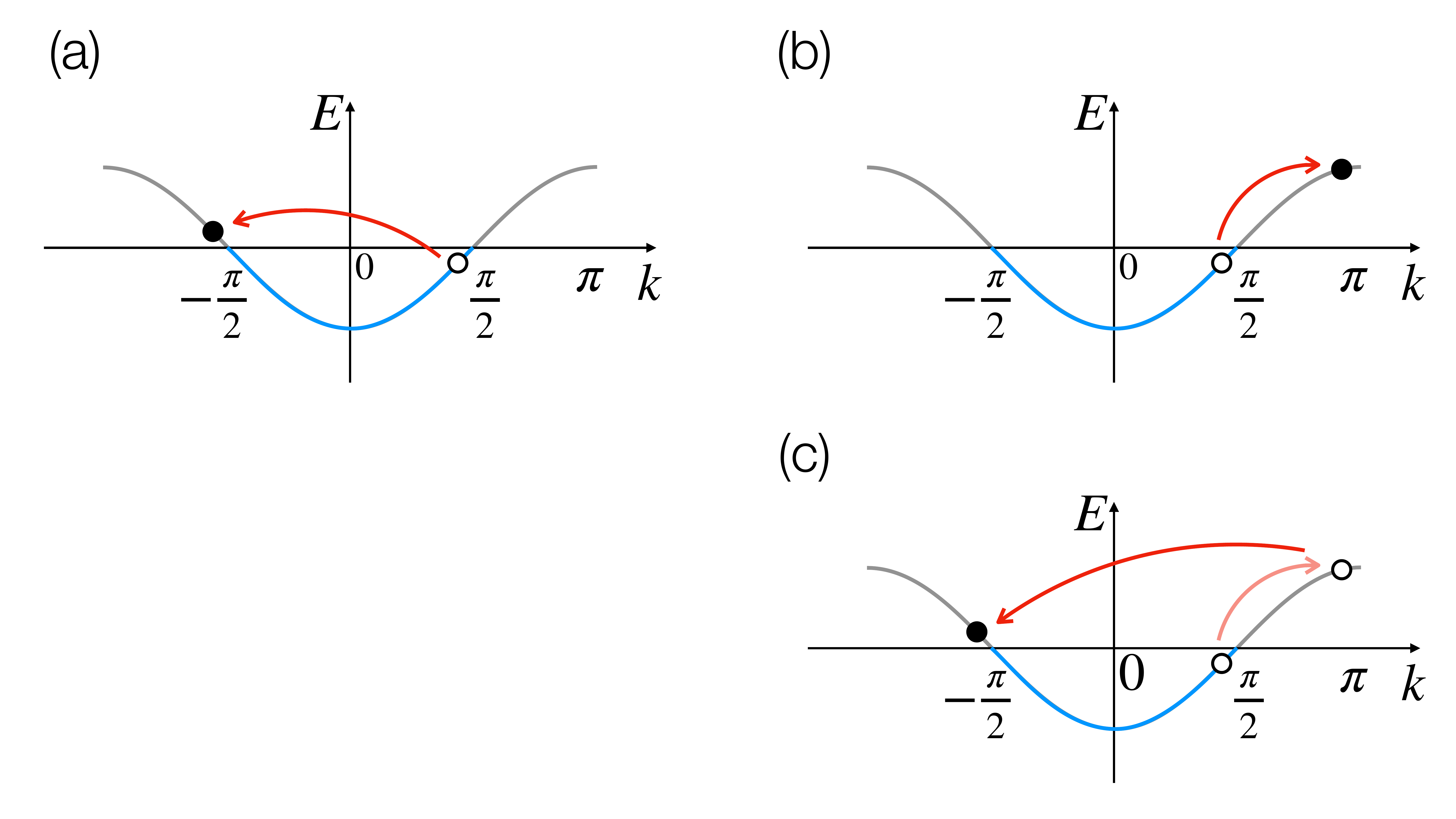}
    \caption{Particle-hole excitations are schematically drawn. Filled and empty circles depict a particle and a hole, respectively. (a) The twofold staggered field $h_2$ and the bond alternation generate low-energy particle-hole excitations with a wave number $q=\pi$. (b) The fourfold screw field $h_4$ generates high-energy excitations with $q=\pi/2$ when they act on the ground state once. (c) The screw field can generate low-energy excitations with $q=\pi$ when they act on the ground state twice.
    }
    \label{fig:hoppings}
\end{figure}

Let us make some observations on effects of the fourfold screw field on the free spinless fermion chain in the TL-liquid phase.
Particle-hole excitations are the fundamental low-energy excitation in the TL-liquid phase.
An operator $\rho_q = \sum_k c_{k+q}^\dag c_k$, which creates a particle-hole excitation with a wave number $q$, can be written as a superposition of bosonic creation and annihilation operators of the TL liquid~\cite{giamarchi_book}.
When the second line of Eq.~\eqref{H_XY_k} acts on the ground state of the XY model, particle-hole excitations with wave numbers $q=\pm \pi/2$ are generated [Fig.~\ref{fig:hoppings}~(b)].
Apparently, such an $h_4$ term hardly affects the low-energy physics of the XY model because these particle-hole excitations have large excitation energies of $O(J)$.
However, applying the $h_4$ term twice to the ground state, I can generate low-energy particle-hole excitations with the wave number $q=\pi$ [Figs.~\ref{fig:hoppings}~(a,c)].
These observations show that though the fourfold screw field is highly irrelevant, it will generates a relevant interaction in a second-order perturbation process.

\subsection{Low-energy effective Hamiltonian}\label{sec:H_eff}

To confirm the perturbative generation of the relevant interaction, I derive a simple effective Hamiltonian that governs the low-energy physics of the fermion chain \eqref{H_XY_k}.
Among several options to derive such a low-energy effective Hamiltonian is
to use the Schrieffer-Wolff canonical transformation~\cite{schriefferwolff, kim_deg}.
The generic theory is explained in Appendix~\ref{app:H_eff_gen}.
Here, I briefly summarize the derivation.
First, I perform a canonical transformation,
\begin{align}
    \mathcal H'_{\rm XY} := e^{\eta} \mathcal H_{\rm XY} e^{-\eta}    
    \label{canonical_transf}
\end{align}
with an antiunitary operator $\eta$.
Two Hamiltonians $\mathcal H_{\rm XY}$ and $\mathcal H'_{\rm XY}$ have one-to-one corresponding lists of eigenstates with exactly the same eigenenergies.
Next, I perform a perturbative expanison by using a projection operator $P$ onto a low-energy subspace $\{\ket{\phi_k}\}_k$ with $k \in R$ and
\begin{align}
    R = \bigl\{ k \in (-\pi, \pi]| \, 0\le \bigl||k|-|k_F|\bigr| < \Lambda \bigr\},
    \label{R}
\end{align}
where $\Lambda $ is a cutoff in the wave number and assumed as $\Lambda \ll 1$.
Here, $\ket{\phi_k}$ is an eigenstate of the unperturbed Hamiltonian, Eq.~\eqref{H_XY_k} for $h_4=0$, with the total wave number $k$.
$P$ can explicitly be written as $P = \sum_{k\in R} \ket{\phi_k}\bra{\phi_k}$.
The projection onto the low-energy subspace leads to the effective Hamiltonian
\begin{align}
    \tilde{\mathcal H}_{\rm XY} &= P \mathcal H'_{\rm XY} P.
    \label{H_eff_def}
\end{align}
Choosing $\eta$ properly, I can simplify the perturbative expansion of the right hand side of Eq.~\eqref{H_eff_def}.
The effective Hamiltonian up to the second order of $h_4/J$ is then given by (see Appendix~\ref{app:H_eff_fermion})
\begin{align}
    &\tilde{\mathcal H}_{\rm XY}
    \notag \\
    &= \sum_{k\in R} \epsilon(k)  c_k^\dag c_k
    -i\frac{h_4^2}{4}\sum_{k\in R} (c_{k+\pi}^\dag c_k - c_{k}^\dag c_{k+\pi}) 
    \notag \\
    &\qquad \times \biggl( \frac{1}{\epsilon(k) - \epsilon(k+\frac{\pi}{2})} + \frac{1}{\epsilon(k+\pi) - \epsilon(k+\frac \pi 2)} \biggr)
    \notag \\
    &\quad + \frac{h_4^2}{2} \sum_{k\in R} c_k^\dag c_k  \biggl( \frac{1}{\epsilon(k) - \epsilon(k-\frac \pi 2)} + \frac{1}{\epsilon(k) -\epsilon(k+\frac \pi 2)} \biggr).
    \label{H_eff_XY}
\end{align}
Since the cutoff $\Lambda$ is small enough, the kinetic term of Eq.~\eqref{H_eff_XY} can be linearized around the Fermi surface~\cite{giamarchi_book}.
Creation operators $c_k^\dag$ at $k\approx \pi/2$ and $k\approx -\pi/2$ are replaced to those of different species, which I denote as $c_{k,R}^\dag$ and $c_{k,L}^\dag$, respectively.
$R$ and $L$ refer to right movers and left movers of fermions.
The low-energy Hamiltonian thus turns out to be
\begin{align}
    \tilde{\mathcal H}_{\rm XY}
    &\approx \sum_{k\in R} \{ v_F(k-k_F) c_{k,R}^\dag c_{k,R} - v_F(k+k_F) c_{k,L}^\dag c_{k,L} \}
    \notag \\
    &\qquad + i\frac{h_4^2}{J} \sum_{k\in R} (c_{k,R}^\dag c_{k,L} - c_{k,L}^\dag c_{k,R}),
    \label{H_eff_XY_linear}
\end{align}
where $v_F$ is the Fermi velocity.
Note that the last term of Eq.~\eqref{H_eff_XY} was discarded in the linearized Hamiltonian \eqref{H_eff_XY_linear} because the coefficient of $c_k^\dag c_k$ is $O(\Lambda h_4^2/J)$ and thus negligibly small for $k\in R$.
The Hamiltonian \eqref{H_eff_XY_linear} is diagonalized in terms of Majorana fermions,
\begin{align}
    \xi_{k,\nu} &= \frac{c_{k,\nu} + c_{k,\nu}^\dag}{\sqrt 2}, \quad
    \chi_{k,\nu} = \frac{c_{k,\nu} - c_{k,\nu}^\dag}{\sqrt 2 i}
\end{align}
The second line of Eq.~\eqref{H_eff_XY_linear} becomes mass terms,
\begin{align}
    i\frac{h_4^2}{J} \sum_{k\in R} (c_{k,R}^\dag c_{k,L} - c_{k,L}^\dag c_{k,R})
    &=i \frac{h_4^2}{J} \sum_{k\in R} (\xi_R \xi_L + \chi_R\chi_L),
    \label{h4_fermion}
\end{align}
which indicate that these Majorana fermions have the excitation gap $\Delta = h_4^2/J$~\cite{shelton_ladder}.
Note that the gap does not reproduce the power law \eqref{gap_num}.
This discrepancy of the power is attributed to interactions of fermions.
I will come back to this point in Sec.~\ref{sec:bosonization}.

The mass terms \eqref{h4_fermion} are nothing but the bond alternation $J\delta_\perp \sum_j (-1)^j  (S_j^x S_{j+1}^x + S_j^y S_{j+1}^y)$ with $\delta_\perp = (h_4/J)^2$~\cite{giamarchi_book}:
\begin{align}
    \sum_j &(-1)^j (S_j^x S_{j+1}^x + S_j^y S_{j+1}^y)
    \notag \\
    &= -\frac{1}{2} \sum_j (-1)^j (c_j^\dag c_{j+1} + c_{j+1}^\dag c_j)
    \notag \\
    &\approx i\sum_{k\in R}(c_{k,R}^\dag c_{k,L} - c_{k,L}^\dag c_{k,R}).
    \label{Dperp_fermion}
\end{align}
For comparison, the following is the twofold staggered field term in terms of fermions.
\begin{align}
    h_2\sum_j (-1)^j S_j^z \approx h_2 \sum_{k\in R} (c_{k,R}^\dag c_{k,L} + c_{k,L}^\dag  c_{k,R}).
    \label{h2_fermion}
\end{align}

Here, I comment on effects of the canonical transformation \eqref{canonical_transf} on observables.
The canonical transformation also transforms an operator, say, $\mathcal O$, in the original model $\mathcal H_{\rm XY}$ to
\begin{align}
    \tilde{\mathcal O}
    &= e^{\eta} \mathcal O e^{-\eta}.
\end{align}
Note that one observes $\braket{\mathcal O}$ in experiments, not $\braket{\tilde{\mathcal O}}$.
Equations~\eqref{Dperp_fermion} and \eqref{h2_fermion} refer to the latter.
According to the generic framework of Appendix~\ref{app:H_eff_gen}, the operator $e^{\eta}$ can be expanded with $h_4/J$,
\begin{align}
    \tilde{\mathcal O} = \mathcal O + [\eta, \mathcal O] + [\eta, \mathcal O] + \frac 12 [\eta, [\eta, \mathcal O]] + \cdots
    \label{tO2O}.
\end{align}
A relation $\braket{\tilde{\mathcal O}} \approx \braket{\mathcal O}$ is valid for $h_4/J \ll 1$.
I can thus basically identify $\mathcal O$ and $\tilde{\mathcal O}$ but their small discrepancy, $[\eta_1, \mathcal O]$, would affect dynamics of spin chains (see Appendix.~\ref{app:esr}).

\subsection{Symmetries}

I showed that the fourfold screw field yields the bond alternation instead of the twofold staggered field.
Actually, the bond-centered inversion symmetry of the spin chain \eqref{H4} forbids the twofold staggered field from emerging in the effective Hamiltonian \eqref{H_eff_XY_linear}.

The uniform spin chain is symmetric under two types of spatial inversions: the site-centered inversion $I_s$ and the bond-centered inversion $I_b$.
These spatial inversions act on spins as
$I_s : \bm S_j \mapsto \bm S_{-j}$ and $I_b: \bm S_j \mapsto \bm S_{1-j}$.
The twofold staggered field is invariant under $I_s$ but not under $I_b$.
On the other hand, the bond alternation and the fourfold screw field are invariant under $I_b$ but not under $I_s$.
In general, a low-energy effective Hamiltonian keeps the symmetries that the original Hamiltonian possesses.
In this sense, the low-energy effective Hamiltonian of the spin chain \eqref{H4} cannot have the twofold staggered field term that breaks the bond-centered inversion symmetry of the original Hamiltonian \eqref{H_XY}.

\section{Interacting boson theory}\label{sec:bosonization}

The second-order perturbation turned out to give rise to the bond alternation in the low-energy effective Hamiltonian of the XY model in the screw field \eqref{H_XY}.
However, the free spinless fermion theory does not explain the power-law behavior of the excitation gap.
In this section, I present a simple theoretical explanation for the numerically found power law, incorporating the interaction of spinless fermions.

Discussions in the previous section prompt us to make an ansatz that the low-energy effective Hamiltonian of the HAFM model in the fourfold screw field \eqref{H4} should be
\begin{align}
    \tilde{\mathcal H}_4
    &:= P e^{\eta} \mathcal H_4 e^{-\eta} P
    \notag \\
    &= J \sum_j \bm S_j \cdot \bm S_{j+1} \notag \\
    &\qquad + J \delta_\perp \sum_j (-1)^j (S_j^x S_{j+1}^x + S_j^y S_{j+1}^y ).
    \label{H_4_eff}
\end{align}
Here, the effective bond alternation is characterized by the parameter  $\delta_\perp \propto (h_4/J)^2$.

Let us investigate whether the ansatz \eqref{H_4_eff} explains numerical results.
For small enough $h_4/J$, one can bosonize the spin operator~\cite{affleck_nonabelian},
\begin{align}
    S_j^z &= \frac{1}{\sqrt{2\pi}} \partial_x \phi + (-1)^j a_1 \sin (\sqrt{2\pi} \phi), \\
    S_j^+ &= e^{-i\sqrt{2\pi}\theta} \bigl[ (-1)^j b_0 + b_1 \sin (\sqrt{2\pi}\phi) \bigr],
\end{align}
where $S_j^+ = S_j^x + i S_j^y$ is the ladder operator.
Coefficients $a_1$, $b_0$, $b_1$ depend on details of the lattice model and are thus nonuniversal.
They are numerically estimated~\cite{hikihara_xxz}.
The Hamiltonian is then bosonized as
\begin{align}
    \tilde{\mathcal H}_4
    &= \frac v{2} \int dx \bigl\{(\partial_x \theta)^2 + (\partial_x\phi)^2 \bigr\}
    \notag \\
    &\qquad + d_{xy} J\delta_\perp \int dx \cos(\sqrt{2\pi}\phi).
    \label{H_4_boson}
\end{align}
Here, $v$ is the spinon velocity and the coefficient $d_{xy}$ is a nonuniversal constant~\cite{takayoshi_coeff, hikihara_coeff_2017}.
This bosonic field theory \eqref{H_4_boson} is interacting but, fortunately, integrable~\cite{essler_review}.

The lowest-energy excitation gap of the sine-Gordon model \eqref{H_4_boson} is exactly given by~\cite{lukyanov_mass, zamolodchikov_mass, essler_cubenzoate}
\begin{align}
    \Delta &= \frac{2v}{\sqrt{\pi}}\frac{\Gamma(1/6)}{\Gamma(2/3)} \biggl( \frac{d_{xy} \pi J}{2v}\frac{\Gamma(3/4)}{\Gamma(1/4)} \delta_\perp \biggr)^{2/3}.
    \label{Delta}
\end{align}
I thus find
\begin{align}
    \Delta \propto \delta_\perp^{2/3} \propto \biggl(\frac{h_4}{J}\biggr)^{4/3}.
    \label{gap_4/3}
\end{align}
The power $4/3$ shows an excellent agreement with the numerical estimation \eqref{gap_num}. 

The sine-Gordon theory also explains the power-law behavior of the transverse dimer order \eqref{D_perp_num}.
In terms of the sine-Gordon theory, the transverse dimer order is an average of the vertex operator~\cite{lukyanov_mass},
\begin{align}
    D_\perp
    &=d_{xy} \braket{\cos(\sqrt{2\pi}\phi)}
    \notag \\
    &= d_{xy} \biggl[\frac{\Delta \sqrt{\pi}\Gamma(2/3)}{v\Gamma(1/6)}\biggr]^{1/2}
    \exp\biggl[\int_0^\infty \frac{dt}{t}\biggl\{ -\frac 12 e^{-2t} 
    \notag \\
    &\qquad +\frac{\sinh^2 (t/2)}{2\sinh (t/4)\sinh t \cosh (3t/4)} \biggr\}\biggr].
\end{align}
It immediately follows from Eq.~\eqref{gap_4/3} that
\begin{align}
    D_\perp \propto \Delta^{1/2} \propto \biggl(\frac{h_4}{J}\biggr)^{2/3}.
    \label{xdimer_2/3}
\end{align}
The power $2/3$ also agrees excellenetly with the numerical estimation \eqref{D_perp_num}.

The bosonization approach predicts the same power law for the longitudinal dimer order.
When I naively bosonizes the operator $(-1)^j S_j^zS_{j+1}^z$, I obtain
\begin{align}
    (-1)^j S_j^z S_{j+1}^z &\approx \frac{a_1}{\sqrt{2\pi}} \partial_x  \phi(x) \sin[\sqrt{2\pi}\phi(x+a)].
\end{align}
An operator-product expansion on the right hand side~\cite{cardy_book} yields a more relevant interaction~\cite{starykh_checkerboard},
\begin{align}
    (-1)^j S_j^z S_{j+1}^z &\approx d_z \cos[\sqrt{2\pi}\phi(x)]
    \notag \\
    &\qquad + \frac{aa_1}{\sqrt{2\pi}} \partial_x\phi(x) \sin[\sqrt{2\pi}\phi(x)] + \cdots.
    \label{D_parallel_boson}
\end{align}
Here, $d_z$ is a nonuniversal constant and precisely estimated~\cite{takayoshi_coeff, hikihara_coeff_2017}.
Note that $d_z$ and $d_{xy}$ satisfy the following relation for small $h_4/J$~\cite{takayoshi_coeff, hikihara_coeff_2017},
\begin{align}
    2d_z &= d_{xy},
    \label{2dz=dxy}
\end{align}
which reflects the SU(2) symmetry of the exchange interaction.
The bosonization formula \eqref{D_parallel_boson} indicates 
\begin{align}
    D_\parallel &= d_z \braket{\cos(\sqrt{2\pi}\phi)}
    \propto \biggl(\frac{h_4}{J}\biggr)^{2/3}.
\end{align}
Nevertheless, the DMRG result (Fig.~\ref{fig:zdimer-vs-hb}) implies Eq.~\eqref{D_parallel_num}.
This discrepancy remains unclear unfortunately.
This will be because the low-energy Hamiltonian fails to capture the uniform $f^z$ order \eqref{fz} properly.

\begin{figure}[t!]
    \centering
    \includegraphics[viewport = 0 0 864 504, width=\linewidth]{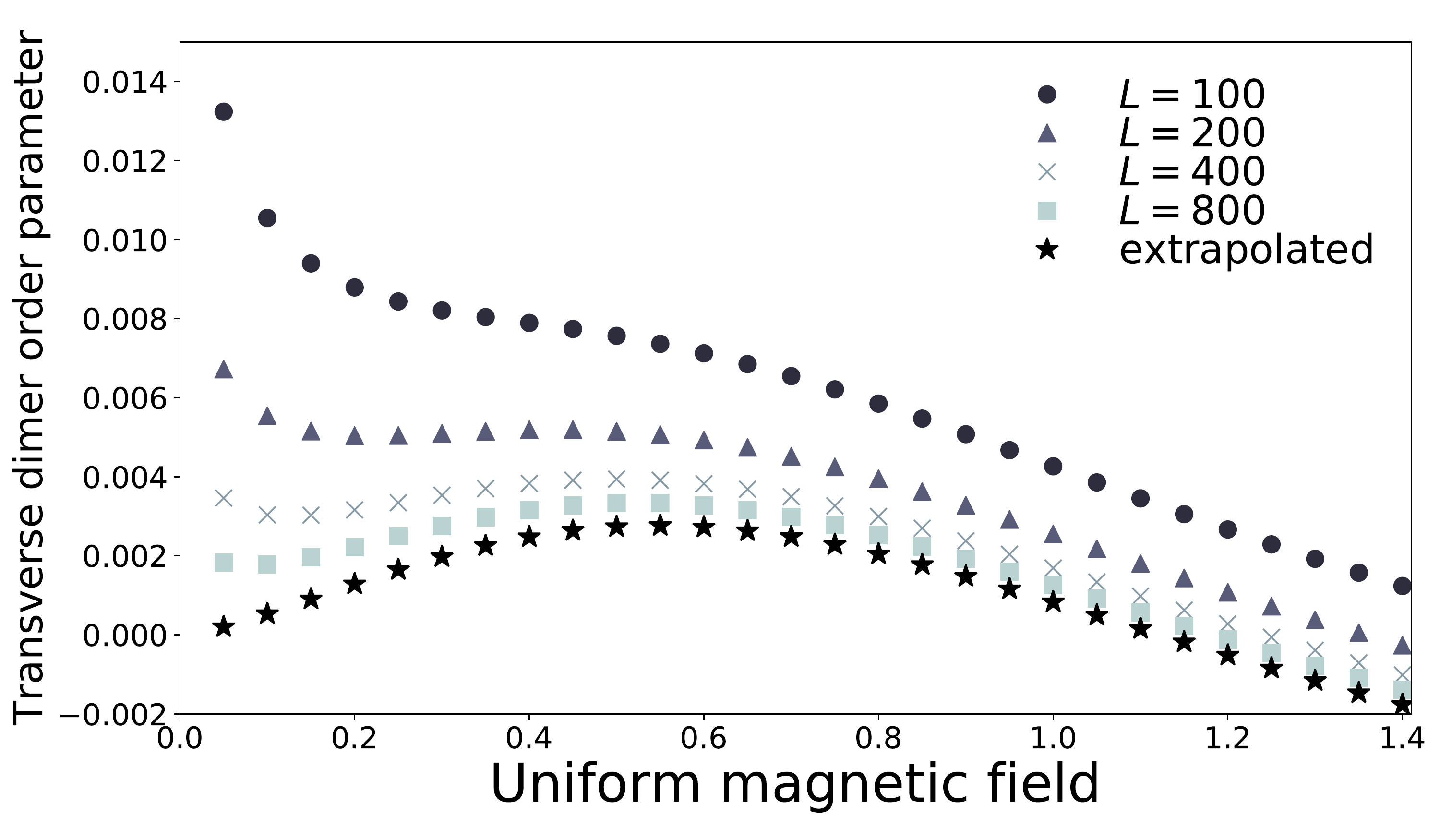}
    \caption{The transverse dimer order parameter \eqref{D_perp} for system sizes $L=100, 200, 400, 800$ and its extrapolated value to the $L \to + \infty$ limited are plotted. 
    The error in the extrapolation is estimated as $< 1\times 10^{-3}~\%$.
    The twofold staggered field $h_2=0.8h_0$ and the screw field $h_4=0.4h_0$ are increased linearly with the uniform magnetic field $h_0$.}
    \label{fig:xydimer-vs-h0}
\end{figure}
\begin{figure}[t!]
    \centering
    \includegraphics[viewport = 0 0 864 504, width=\linewidth]{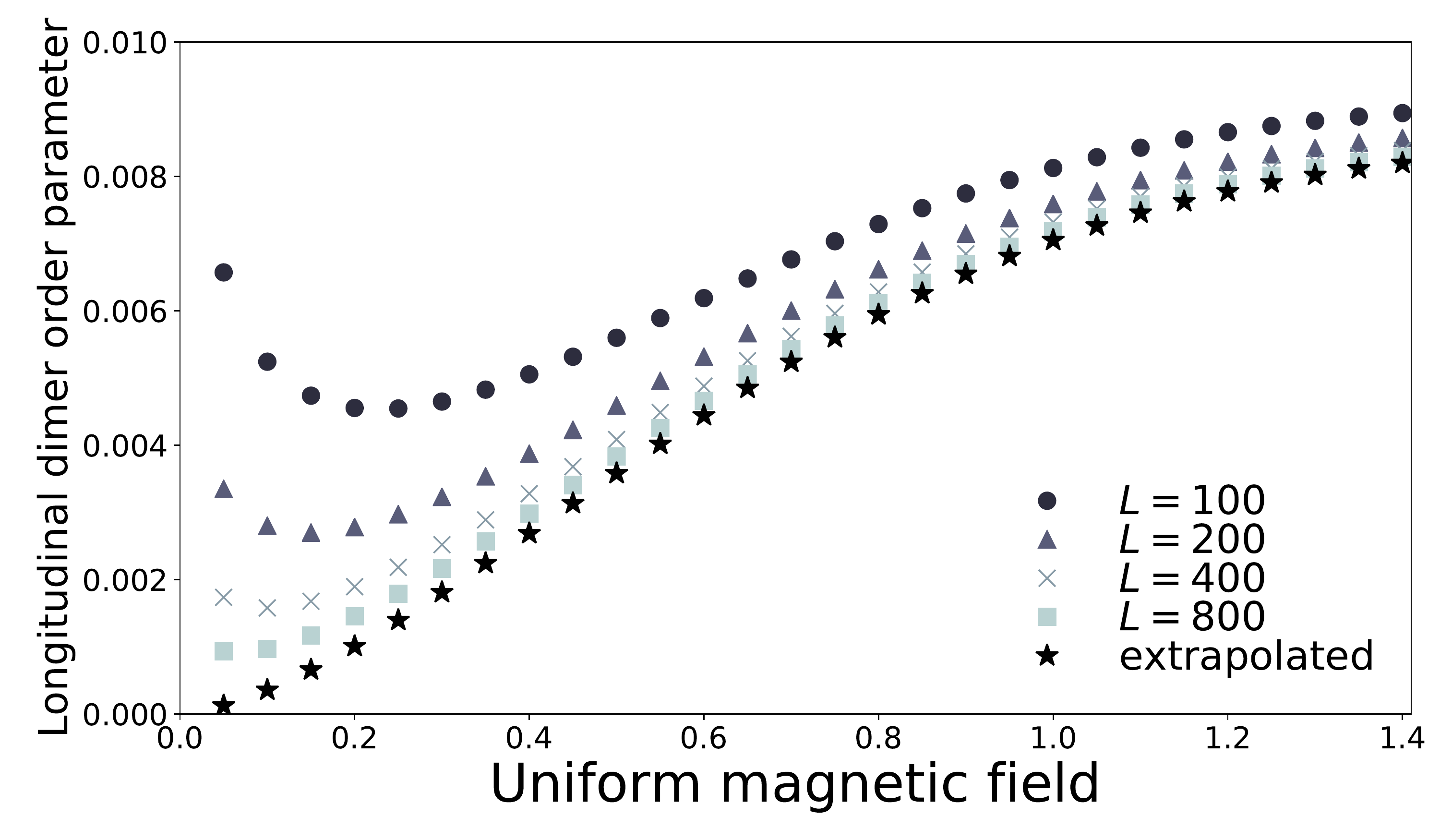}
    \caption{The longitudinal dimer order parameter \eqref{D_parallel} for system sizes $L=100, 200, 400, 800$ and its extrapolated value to $L \to + \infty$ are plotted.
    The error in the extrapolation is estimated as $< 1\times 10^{-3}~\%$.
    Parameters $(\alpha_2, \alpha_4) = (0.8, 0.4)$ are used.}
    \label{fig:zdimer-vs-h0}
\end{figure}

\begin{figure}
    \centering
    \includegraphics[viewport = 0 0 864 504, width=\linewidth]{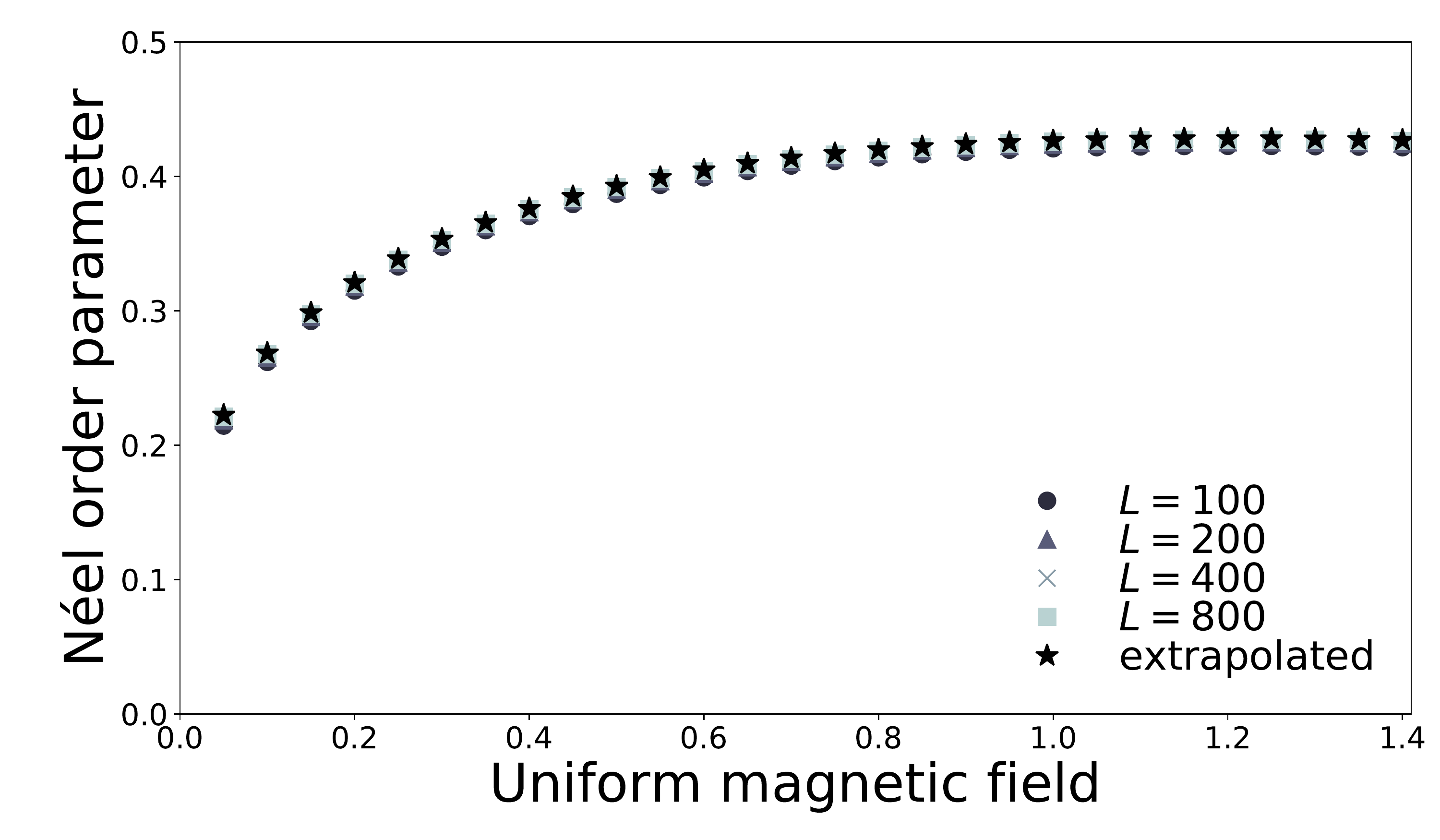}
    \caption{The N\'eel order parameter \eqref{Nx} for  system sizes $L=100, 200, 400, 800$, and its extrapolated value to $L \to + \infty$ are plotted.
    The error in the extrapolation is estimated as $< 1\times 10^{-6}~\%$.
    Parameters $(\alpha_2, \alpha_4) = (0.8, 0.4)$ are used.
    }
    \label{fig:neel-vs-h0}
\end{figure}

\section{Coexistence of N\'eel and dimer orders}\label{sec:coex}

\subsection{Renormalization groups}\label{sec:rg}

On the basis of the fact that the fourfold screw field induces the transverse dimer order \eqref{xdimer_2/3}, here, I investigate the realistic case with $h_2$ and $h_4$ proportional to the uniform field $h_0$.
I assume two proportional coefficients 
\begin{align}
    \alpha_2 = h_2/h_0, \qquad
    \alpha_4 = h_4/h_0,
    \label{coeff}
\end{align}
are both constant.
DMRG results for the dimer order parameters and the N\'eel order parameter are shown in Figs.~\ref{fig:xydimer-vs-h0}, \ref{fig:zdimer-vs-h0}, and \ref{fig:neel-vs-h0} for $(\alpha_2, \alpha_4) = (0.8, 0.4)$.

To understand DMRG results, I replace the Hamiltonian \eqref{H_def} to the low-energy effective Hamiltonian,
\begin{align}
    \tilde{\mathcal H}
    &= J \sum_j \bm S_j \cdot \bm S_{j+1} - h_0 \sum_j S_j^z - h_2 \sum_j (-1)^j S_j^x
    \notag \\
    &\qquad +J \delta_\perp \sum_j (-1)^j (S_j^x S_{j+1}^x + S_j^y S_{j+1}^y )
    \label{H_main}
\end{align}
with $\delta_\perp \propto (h_4/J)^2$.
I can immediately bosonize it.
\begin{align}
    \tilde{\mathcal H}
    &= \frac{v}{2} \int dx \bigl\{(\partial_x \theta)^2 + (\partial_x \phi)^2 \bigr\}
    -\frac{h_0}{\sqrt{2\pi}} \int dx \, \partial_x \phi
    \notag \\
    &\qquad -g_2 \int dx \, \cos(\sqrt{2\pi} \theta) + g_4 \int dx \, \cos(\sqrt{2\pi}\phi),
    \label{H_bosonized}
\end{align}
where $g_2=  b_0 h_2$ and $g_4 = d_{xy} J \delta_\perp$.
This complex Hamiltonian consists of two parts.
The first line of Eq.~\eqref{H_bosonized} favors the gapless TL-liquid ground state for small $h_0/J$.
The second line represents potential terms of $\phi$ and $\theta$ that give rise to an excitation gap.
In general, the scaling dimensions of $\cos(\sqrt{2\pi}\theta)$ and $\cos(\sqrt{2\pi}\phi)$ are  $1/4K$ and $K$, respectively.
Here, $K$ is a parameter called the Luttinger parameter that signifies strength of interactions~\cite{giamarchi_book}.
The XY and the Heisenberg chains have $K=1$ and $K=1/2$, respectively.
In the latter case, two cosine interactions are equally relevant.
Therefore, N\'eel and dimer orders can coexist in the ground state from the viewpoint of the renormalization group (RG).

The coupling constant $g_2$ of $\cos(\sqrt{2\pi}\theta)$, whose bare value is $b_0 h_2$, is increasing in the course of iterative RG transformations.
$g_2$ follows the RG equation, 
\begin{align}
    \frac{dg_2(\ell)}{d\ell} &\approx \frac 32 g_2(\ell).
    \label{RG_g2}
\end{align}
Here, $\ell$ characterizes the effective short-distance cutoff $a(\ell) = a e^{\ell}$.
Note that $a$ is the lattice spacing which was set to be unity.
The RG transformation of Eq.~\eqref{RG_g2} is terminated when $a(\ell)$ reaches a correlation length of the lowest-energy excitation, $v/\Delta$.

Despite the same value of scaling dimensions, behaviors of the RG transformation of $g_4$ differ from that of $g_2$.
This is due to the Zeeman energy which competes with the transverse bond alternation $\cos(\sqrt{2\pi}\phi)$.
I can absorb the Zeeman energy in Eq.~\eqref{H_bosonized} into the kinetic term by shifting $\phi \to \phi + h_0/\sqrt{2\pi} v$.
The $\phi$ shift introduces an incommensurate oscillation to the transverse bond alternation term, $\cos(\sqrt{2\pi}\phi) \to \cos(\sqrt{2\pi}\phi + h_0x/v)$.
When the wave length $v/h_0$ is much longer than the short-distance cutoff $a(\ell)$, the incommensurate oscillation is negligible and then the RG equation of $g_4$ is simply
\begin{align}
    \frac{dg_4(\ell)}{d\ell} &\approx \frac 32 g_4(\ell).
    \label{RG_g4}
\end{align}
Otherwise, the incommensurate oscillation is rapid enough to eliminate $g_4$:
\begin{align}
    g_4(\ell) = 0.
    \label{RG_g4_0}
\end{align}
The same argument can be found in Ref.~\cite{affleck_fig}.

Now I can classify into two cases the strong-coupling limit that the RG flow eventually reaches.
(i) When $v/\Delta \ll v/h_0$, the coupling constants $g_2(\ell)$ and $g_4(\ell)$ grow equally following Eqs.~\eqref{RG_g2} and \eqref{RG_g4} and eventually reach $O(1)$.
Then the N\'eel and the dimer orders coexist in the ground state.
(ii) When $v/\Delta \gg v/h_0$, the coupling constant $g_4(\ell)$ vanishes because of the rapid incommensurate oscillation [Eq.~\eqref{RG_g4_0}].
Then, the ground state has only the N\'eel order.

The correlation length $v/\Delta$ and the wave length $v/h_0$ are easily compared.
If $\alpha_2/\alpha_4 =0$, the gap becomes $\Delta \propto (h_4/J)^{2/3} \propto (h_0/J)^{4/3}$.
When $h_0/J\ll 1$, the gap $\Delta \propto (h_0/J)^{4/3}$ never exceeds $h_0/J$, in other words, $v/\Delta \gg v/h_0$.
Then the ground state does not have the dimer order.
On the other hand, if $\alpha_2/\alpha_4$ is finite, the gap is a complex function of $h_0/J$.
Still, in the limit $h_0/J \to 0$, the gap is reduced to the simple form of $\Delta \propto (h_0/J)^{2/3}$, which is much larger than $h_0/J$.
In other words, $v/\Delta \ll v/h_0$ is valid and the first scenario comes true.
Therefore, finite $|\alpha_2/\alpha_4|$ is necessary for the coexistence of the N\'eel and the dimer orders.

\subsection{Non-Abelian bosonization}\label{sec:nonAbelian}

There is one remaining problem in the RG analysis on the coexistence of the N\'eel and the dimer orders.
The transverse N\'eel order $(-1)^j S_j^x \approx \cos(\sqrt{2\pi} \theta)$ and the transverse dimer order $\cos(\sqrt{2\pi}\phi)$ seem to compete with each other since $\phi$ and $\theta$ are noncommutative.
However, this competition is an artifact of the Abelian bosonization and these orders are cooperative~\cite{garate_dm, chan_dm, jin_dm}.

To avoid the artifact, I rewrite the Hamiltonian \eqref{H_main} as
\begin{align}
    \tilde{\mathcal H}
    &= J \sum_j \biggl\{1+\frac{2\delta_\perp}3 (-1)^j \biggr\} \bm S_j \cdot \bm S_{j+1} -h_2 \sum_j (-1)^j S_j^x
    \notag \\
    &\qquad -\frac{J\delta_\perp}{3} \sum_j (-1)^j(2S_j^z S_{j+1}^z - S_j^x S_{j+1}^x - S_j^y S_{j+1}^y),
    \label{H4_eff_su2}
\end{align}
where I assume finite $\alpha_2/\alpha_4$.
According to the RG analysis, the uniform Zeeman energy is negligible for finite $\alpha_2/\alpha_4$.
Here, I simply put $h_0=0$ from the beginning.
Note that the second line of Eq.~\eqref{H4_eff_su2} yields only irrelevant interactions for the relation \eqref{2dz=dxy} and is discarded hereafter.

Instead of the Abelian bosonization, I employ the non-Abelian bosonization approach~\cite{affleck_nonabelian, gogolin_book}.
In the non-Abelian bosonization language, the effective Hamiltonian \eqref{H_bosonized} for $h_0=0$ is written as
\begin{align}
    \tilde{\mathcal H}
    &= \frac{2\pi v}{3} \int dx \,(\bm J_R \cdot \bm J_R +\bm J_L \cdot \bm J_L)
    \notag \\
    &\qquad +\frac{d_{xy} J\delta_\perp}3 \int dx \, \tr(g) -i\frac{b_0 h_2}2 \int dx \, \tr (g \sigma^x).
    \label{H_nonabelian}
\end{align}
Here, the spin operator $\bm S_j$ is represented as
\begin{align}
    \bm S_j = \bm J_R + \bm J_L  -\frac{ib_0}2 \tr(g\bm \sigma),
    \label{S2g}
\end{align}
with $\mathrm{SU(2)}$ currents $\bm J_R$ and $\bm J_L$, a fundamental field $g\in \mathrm{SU(2)}$, the Pauli matrices $\bm \sigma = (\sigma^x \, \sigma^y \, \sigma^z)^T$~\cite{affleck_nonabelian, gogolin_book}.
The matrix $g \in \mathrm{SU(2)}$ is simply related to the U(1) bosons,
\begin{align}
    g &= 
    \begin{pmatrix}
    e^{i\sqrt{2\pi}\phi} & ie^{-i\sqrt{2\pi}\theta} \\
    ie^{i\sqrt{2\pi}\theta} & e^{-i\sqrt{2\pi}\phi}
    \end{pmatrix}.
    \label{g_matrix}
\end{align}
Since global rotations keep the excitation spectrum unchanged, I perform a global $\pi/2$ rotation in the spin space as $(\sigma^x, \, \sigma^y, \, \sigma^z) \to (\sigma^z, \, \sigma^y, \, -\sigma^x)$.
The rotation transforms the Hamiltonian \eqref{H_nonabelian} into
\begin{align}
    \tilde{\mathcal H}
    &= \frac{2\pi v}{3} \int dx \,(\bm J_R \cdot \bm J_R +\bm J_L \cdot \bm J_L)
    \notag \\
    &\qquad +\frac{d_{xy} J\delta_\perp}3 \int dx \, \tr(g) -i\frac{b_0 h_2}2 \int dx \, \tr (g \sigma^z).
\end{align}
Translating it to the Abelian bosonizaion language, I can express this Hamiltonian as
\begin{align}
    \tilde{\mathcal H}
    &= \frac{v}{2} \int dx \bigl\{ (\partial_x\theta)^2 + (\partial_x \phi)^2 \bigr\}
    \notag \\
    &\quad +\frac{2d_{xy} J\delta_\perp}3 \int dx \cos(\sqrt{2\pi}\phi) - b_0 h_2 \int dx \, \sin(\sqrt{2\pi}\phi)
    \notag \\
    &=\frac{v}{2} \int dx \bigl\{ (\partial_x\theta)^2 + (\partial_x \phi)^2 \bigr\} + g \int dx \cos(\sqrt{2\pi}\phi + \alpha),
    \label{H_0h0_abelian}
\end{align}
with the coupling constant $g=\sqrt{(2d_{xy}J\delta_\perp/3)^2 + (b_0h_2)^2}$ and the phase shift $\alpha = \tan^{-1}(3b_0h_2/2d_{xy}J\delta_\perp)$.
The complex model \eqref{H_main} is thus reduced to the simple sine-Gordon model \eqref{H_0h0_abelian}.

The incommensurate phase shift $\alpha$ realizes the coexistence of the N\'eel order, 
\begin{align}
    N_x := \sum_j (-1)^j\braket{S_j^x}/L, 
    \label{Nx}
\end{align}
and the transverse dimer order \eqref{D_perp}.
Their ground-state averages are given by
\begin{align}
    N_x
    &\propto \sqrt{\Delta/J} \sin \alpha \propto (g/J)^{1/3}\sin \alpha,
    \label{Nx_SG} \\
    D_\perp
    &\propto\sqrt{\Delta/J} \cos \alpha \propto  (g/J)^{1/3}\cos \alpha.
    \label{D_perp_SG}
\end{align}
For $h_0/J \to 0$, the angle $\alpha$ approaches $\pi/2$.
Therefore, at low fields $h_0/J\ll 1$, $N_x$ and $D_\perp$ are expected to follow power laws:
\begin{align}
    N_x &\propto (h_0/J)^{1/3}, \quad D_\perp \propto (h_0/J)^{4/3}.
    \label{Nx-Dxy}
\end{align}
The power law \eqref{Nx-Dxy} is qualitatively consistent with Figs.~\ref{fig:xydimer-vs-h0} and \ref{fig:neel-vs-h0} at low fields.
For weak magnetic fields, the longitudinal and the transverse dimer order parameters are almost equal.
On the other hand, they are much smaller than the N\'eel order parameter $N_x$.
If we assume $\delta_\perp = C_\perp {h_4}^2$, the bosonized theory \eqref{H_0h0_abelian} predicts a ratio  $D_\perp/N_x$ given by
\begin{align}
    \frac{D_\perp}{N_x}&= \frac{d_{xy}}{b_0} \cot \alpha\notag \\
    &= \frac{2C_\perp}3 \biggl(\frac{d_{xy}}{b_0}\biggr)^2\frac{{\alpha_4}^2}{\alpha_2} \frac{h_0}{J}.
    \label{ratio}
\end{align}
The right hand side is approximately estimated as $0.028C_\perp h_0/J$ for the parameters used in DMRG.
If $C_\perp \approx 0.7$, the ratio \eqref{ratio} is roughly consistent with the DMRG data of Figs.~\ref{fig:xydimer-vs-h0} and \ref{fig:neel-vs-h0} for $h_0/J< 0.3$.

\section{Experimental relevance}\label{sec:cupm}

\subsection{[Cu(pym)(H$_2$O)$_4$]SiF$_6 \cdot$ H$_2$O}

The model \eqref{H_def} that I have dealt with so far is similar to a model proposed for the spin-chain compound \cupm{} with the following Hamiltonian:~\cite{liu_chain_esr}
\begin{align}
    \mathcal H_{\rm exp}
    &= J \sum_j (S_j^x S_{j+1}^x + S_j^y S_{j+1}^y + \lambda S_j^z S_{j+1}^z)
    \notag \\
    &\qquad + h_0 \sum_j S_j^x + h_2 \sum_j (-1)^j S_j^y + h_4 \sum_j \delta_j S_j^z
    \notag \\
    &\qquad + D_u \sum_j (S_j^x S_{j+1}^y - S_j^y S_{j+1}^x),
    \label{H_cupm}
\end{align}
where $\lambda \approx 1$.
There are three differences in two models \eqref{H_def} and \eqref{H_cupm}: field directions, the weak exchange anisotropy, and the uniform DM interaction.
In this section, I investigate effects of these differences one by one and discuss an experimental feasibility of the field-induced transverse dimer order.

\subsection{Field directions}

In the model \eqref{H_cupm}, the magnetic field $h_0$, the twofold staggered field $h_2$, and the fourfold screw field $h_4$ are applied in different directions.
On the other hand, the model \eqref{H_def} has the uniform field and the fourfold screw field in the same direction.
This difference in field directions is actually insignificant in low-energy physics for small $h_0/J$.
When the uniform DM interaction is absent ($D_u=0$), the model \eqref{H_cupm} has the following low-energy effective Hamiltonian:
\begin{align}
    \tilde{\mathcal H}_{\rm exp}
    &\approx J \sum_j (S_j^x S_{j+1}^x + S_j^y S_{j+1}^y + \lambda S_j^z S_{j+1}^z)
    \notag \\
    &\qquad + h_0 \sum_j S_j^x + h_2 \sum_j(-1)^j S_j^y
    \notag \\
    &\qquad + J \delta_\perp \sum_j (-1)^j (S_j^x S_{j+1}^x + S_j^y S_{j+1}^y),
\end{align}
with $\delta_\perp \propto (h_4/J)^2$.
Relabeling the spins $(S_j^x, \, S_j^y, \, S_j^z) \to (S_j^z, \, S_j^x, \, S_j^y)$, I rewrite this Hamiltonian as
\begin{align}
    \tilde{\mathcal H}_{\rm exp}
    &\approx J \sum_j ( S_j^x S_{j+1}^x + \lambda S_j^y S_{j+1}^y +  S_j^z S_{j+1}^z)
    \notag \\
    &\qquad +h_0 \sum_j S_j^z + h_2 \sum_j (-1)^j S_j^x
    \notag \\
    &\qquad + \frac{2J \delta_\perp}3 \sum_j (-1)^j \bm S_j \cdot \bm S_{j+1} 
    \notag \\
    &\qquad - \frac{J\delta_\perp}{3} \sum_j (2S_j^y S_{j+1}^y - S_j^x S_{j+1}^x - S_j^z S_{j+1}^z).
    \label{H_exp_0Du_rot}
\end{align}
Since the last term of Eq.~\eqref{H_exp_0Du_rot} is irrelevant, the bosonized Hamiltonian of Eq.~\eqref{H_exp_0Du_rot} is given by
\begin{align}
    \tilde{\mathcal H}_{\rm exp}
    &= \frac{v}{2} \int dx \bigl\{ (\partial_x \theta)^2 + (\partial_x \phi)^2\bigr\} 
    + \frac{h_0}{\sqrt{2\pi}} \int dx \, \partial_x \phi
    \notag \\
    &\qquad 
    +g_2 \int dx \, \cos (\sqrt{2\pi}\theta)  + g_4\int dx \, \cos(\sqrt{2\pi}\phi)
    \notag \\
    &\qquad + g_a \int dx \, \sin(\sqrt{8\pi}\theta),
    \label{H_exp_0Du_bosonized}
\end{align}
with $g_a \propto J(\lambda-1)$.
Except for the last term that comes from the exchange anisotropy, the Hamiltonian \eqref{H_exp_0Du_bosonized} is identical to the one \eqref{H_bosonized} investigated in Sec.~\ref{sec:coex}.

\subsection{Exchange anisotropy}\label{sec:aniso}

The bosonized effective Hamiltonian \eqref{H_exp_0Du_bosonized} shows that the small exchange anisotropy $\lambda \approx 1$ gives rise to the $\sin(\sqrt{8\pi}\theta)$ interaction. 
Though this interaction iteself can be marginally relevant at most in the RG sense, it is negligible in the presence of the much more relevant interaction $\cos(\sqrt{2\pi} \theta)$.

\subsection{Uniform DM interaction}\label{sec:dm}

After all, the uniform DM interaction is the only significant difference in the models \eqref{H_def} and \eqref{H_cupm}.
The major effect of the uniform DM interaction is a chiral rotation.
Let us resurrect the uniform DM intearction in the rotated effective Hamiltonian \eqref{H_exp_0Du_rot}:
\begin{align}
    \tilde{\mathcal H}_{\rm exp}
    &\approx J \sum_j ( S_j^x S_{j+1}^x + \lambda S_j^y S_{j+1}^y +  S_j^z S_{j+1}^z)
    \notag \\
    &\qquad +h_0 \sum_j S_j^z + h_2 \sum_j(-1)^j S_j^x
    \notag \\
    &\qquad + \frac{2J \delta_\perp}3 \sum_j (-1)^j \bm S_j \cdot \bm S_{j+1} 
    \notag \\
    &\qquad 
    + D_u \sum_j (S_j^z S_{j+1}^x - S_j^x S_{j+1}^z).
    \label{H_exp_rot}
\end{align}
Here, the irrelevant term is already dropped.
The uniform DM interaction itself is bosonized as~\cite{gangadharaiah_dm_wire}
\begin{align}
    D_u \sum_j (S_j^z S_{j+1}^x - S_j^x S_{j+1}^z)
    &\approx \gamma D_u \int dx \, (J_R^y - J_L^y),
    \label{Du_current}
\end{align}
with a nonuniversal constant $\gamma>0$.
I employed the non-Abelian bosonization language (see Sec.~\ref{sec:nonAbelian}).
The right hand side of Eq.~\eqref{Du_current} resembles the Zeeman energy,
\begin{align}
    h_0 \sum_j S_j^z 
    &= h_0 \int dx \, (J_R^z + J_L^z).
    \label{Zeeman_current}
\end{align}
While the Zeeman energy \eqref{Zeeman_current} is non-chiral (i.e. symmetric in the permutation of $R \leftrightarrow L$), but the uniform DM interaction \eqref{Du_current} is chiral.
As far as only either the $R$ part or the $L$ part is concerned, I cannot distinguish the Zeeman energy and the uniform DM interaction.

A chiral rotation can combine the uniform DM interaction \eqref{Du_current} and the Zeeman energy \eqref{Zeeman_current}~\cite{garate_dm, jin_dm, chan_dm}.
\begin{align}
    \bm J_\nu &= \mathcal R(\theta_\nu) \bm M_\nu,
    \label{rot}
\end{align}
for $\nu=R, L$.
Here, the rotation $R(\theta_\nu)$ is defined as
\begin{align}
    R(\theta_\nu)
    &=
    \begin{pmatrix}
    1 & 0 & 0 \\
    0 & \cos \theta_\nu  & \sin \theta_\nu \\
    0 & -\sin \theta_\nu & \cos \theta_\nu
    \end{pmatrix}.
\end{align}
I assume that $h_0$ and $\gamma D_u$ are both positive.
Then, the rotation leads to
\begin{align}
    h_0 (J_R^z + J_L^z) + \gamma D_u (J_R^y - J_L^y)
    &= t_\phi (M_R^z + M_L^z),
\end{align}
with 
\begin{align}
    t_\phi = \sqrt{{h_0}^2 + (\gamma D_u)^2},
\end{align}
if $\theta_R$ and $\theta_L$ take the following values,
\begin{align}
    \theta_R &= \tan^{-1} \biggl( \frac{\gamma D_u}{h_0} \biggr),
    \label{thetaR} \\
    \theta_L &=  -\theta_R.
    \label{thetaL}
\end{align}
The chiral rotation $R(\theta_\nu)$ transforms $g$ into
\begin{align}
    g' = e^{i\sigma^x \theta_L/2} g e^{-i\sigma^x \theta_R/2}.
    \label{g'}
\end{align}

Effects of this chiral rotation on the low-energy Hamiltonian are explained in Appendix.~\ref{app:esr}.
Here, I show results only.
The chirally rotated Hamiltonian then becomes
\begin{align}
    &\tilde{\mathcal H}_{\rm exp}
    \notag \\
    &= \frac{v}{2} \int dx \bigl\{ (\partial_x\Theta)^2 + (\partial_x \Phi)^2 \bigr\}
    + \frac{t_\phi}{\sqrt{2\pi}} \int dx \, \partial_x \Phi
    \notag \\
    &\quad + g'_2 \int dx \, \{\cos(\sqrt{2\pi} \Theta) \cos \theta_R + \cos(\sqrt{2\pi}\Phi) \sin \theta_R \}
    \notag \\
    &\quad +  g'_4 \int dx \, \{\cos (\sqrt{2\pi}\Phi) \cos \theta_R - \cos(\sqrt{2\pi} \Theta) \sin \theta_R\},
    \label{H_exp_rot_boson}
\end{align}
where $g'_2 \propto h_2$ and $g'_4 \propto J\delta_\perp$.
Note that the chiral rotation \eqref{g'} mixes the N\'eel and dimer orders.
Though the right hand side of Eq.~\eqref{H_exp_rot_boson} is complex, its basic structure is the same as that of Eq.~\eqref{H_bosonized} in a sense that coupling constants $g'_2$ and $g'_4$ in the former follow the same RG equations as those for $g_2$ and $g_4$ in the latter.
Following the argument in Sec.~\ref{sec:nonAbelian}, I obtain the N\'eel and dimer orders in the ground state (see Appendix~\ref{app:esr} for details):
\begin{align}
    N_x
    &\propto  (G/J)^{1/3} \sin \alpha', \\
    D_\perp
    &\propto(G/J)^{1/3} \cos \alpha',
\end{align}
where the coupling constant $G$ and the angle $\alpha'$ are defined in Eqs.~\eqref{G} and \eqref{alpha'}.
In analogy with Eq.~\eqref{Nx-Dxy}, I obtain 
\begin{align}
    N_x \propto (h_0/J)^{1/3}, \quad D_\perp \propto (h_0/J)^{4/3},
\end{align}
at low fields $h_0/J\ll 1$.

In short, the uniform DM interaction causes the chiral rotation that mixes the N\'eel and the transverse dimer orders if the following condition is met.
\begin{align}
     (h_0/J)^{2/3} \gg t_\phi/J.
     \label{cond_dm}
\end{align}
When $D_u=0$, the condition \eqref{cond_dm} is trivially satisfied for small $h_0/J$.
However, the inequality \eqref{cond_dm} can be violated at extremely small magnetic fields $h_0/D_u \ll 1$.

Let me comment on effects of the uniform DM interaction on electron spin resonance (ESR).
In one-dimensional quantum spin systems, the uniform DM interaction splits the ESR peak that corresponds to the Zeeman energy [Eqs.~\eqref{Du_current} and \eqref{Zeeman_current}]~\cite{povarov_dm_esr, furuya_dm_esr}.
In some cases, the DM interaction changes selection rules of ESR and yields an additional resonance that occurs at a frequency away from the Zeeman energy~\cite{oshikawa_esr, furuya_bbs, ozerov_dm_esr, zhao_cubenzoate, lou_midgap}.

The experiment~\cite{liu_chain_esr} on \cupm{} found that ESR peaks of this compound exhibit unconventional power-law dependence on the magnetic field.
A part of this unconventional behavior is attributed to the chiral rotation and the complex dependence of the coupling constant on the magnetic field.
A derivation of ESR selection rules is described in Appendix.~\ref{app:esr}.
Here, I simply summarize the result.
Elementary excitations of the sine-Gordon theory are a soliton, an antisoliton, and their bound states, breathers.
Let us represent these excitation gaps by $M$, where $M$ can be the soliton mass or the breather mass.
ESR in the model \eqref{H_exp_rot} occurs when the frequency $\omega$ of the applied microwave satisfies
\begin{align}
    \omega = M,
    \label{w=m}
\end{align}
or
\begin{align}
    \omega = \sqrt{{t_\phi}^2 + M^2}.
    \label{w=sqrt}
\end{align}
These resonance frequencies are close to neither the Zeeman energy nor the typical gap, $\omega \propto (h_0/J)^{2/3}$, in quantum spin chains with the twofold staggered field~\cite{dender_cubenzoate, zvyagin_cupm_2004, umegaki_kcugaf6}.
In particular, the latter resonance frequency \eqref{w=sqrt} approaches $\omega \to \gamma D_u$ in the $h_0 \to 0$ limit~\cite{povarov_dm_esr, furuya_dm_esr}.

\subsection{(Ba/Sr)Co$_2$V$_2$O$_8$}

I can find other quantum spin chain compounds with the fourfold screw symmetry such as \bacovo{}~\cite{faure_bacovo, kimura_bacovo} and \srcovo{}~\cite{wang_srcovo, bera_srcovo}.
Unlike \cupm{}, these compounds have Ising-like exchange interactions.
As I already showed, the SU(2) symmetry of the exchange interaction is essential for the coexistence of the N\'eel and the transverse dimer orders.
The strong enough Ising anisotropy ruins the coexistence and thus makes the fourfold screw field insignificant.
Thus far, most experimental results on these compounds are well understood with models without the fourfold screw field~\cite{faure_bacovo, kimura_bacovo, wang_srcovo, bera_srcovo} though some ESR peaks can be attributed to the presence of the fourfold screw field~\cite{kimura_bacovo}.

\section{Summary}\label{sec:summary}

I discussed the novel type of field-induced gap phenomena, field-induced dimer orders in quantum spin chains.
The fourfold screw field with the bond-centered inversion symmetry introduces perturbatively the effective bond alternation to the spin chain.
In analogy with the twofold staggered field, the fourfold screw field, which breaks the one-site translation symmetry, gives rise to the excitation gap from the ground state to the excited states.

In the first part of the paper, I applied the fourfold screw field $h_4$ solely to quantum spin chains.
The field-induced excitation gap by $h_4$ turned out to show a distinctive power law from that by the twofold staggered field $h_2$.
The gap is proportional to  $(h_4/J)^{4/3}$ for the fourfold screw field instead of $(h_2/J)^{2/3}$ for the twofold staggered field $h_2$~\cite{oshikawa_stag_prl, affleck_fig}.
The power law was predicted from the quantum field theory and consistent with the numerical results (Fig.~\ref{fig:gap-vs-hb}).
The field theory also gave the explanation on the power law of the transverse dimer order \eqref{D_perp_num}, though it failed for the longitudinal one \eqref{D_parallel_num} somehow.

Next, I applied the uniform field, the twofold staggered field, and the fourfold screw field simultaneously to HAFM chains.
The SU(2) symmetry of the exchange interaction turned out to make the coexistence of the N\'eel and dimer order possible in the ground state.
The coexistence of these orders are nontrivial and already interesting~\cite{garate_dm, jin_dm, chan_dm}.
More interestingly, the dimer order grows in association with the uniform magnetic field (Figs.~\ref{fig:xydimer-vs-h0}, \ref{fig:zdimer-vs-h0}, and \ref{fig:neel-vs-h0}).
The coexistent growth of the N\'eel and the dimer orders were numerically found and supported by the effective field theory.

Last but not least, I discussed the relevance of my model to experimental studies, in particular, Ref.~\cite{liu_chain_esr} on \cupm{}.
There are three differences in the model for \cupm{} and the model Hamiltonian \eqref{H_def}, or equivalently Eq.~\eqref{H_main}, that I dealt with in this paper.
They are field directions, the weak exchange anisotropy, and the uniform DM interaction.
In Sec.~\ref{sec:cupm}, I discussed that all the three differences do not interfere with the field-induced growth of the N\'eel and the dimer orders.
However, the uniform DM interaction may cause nontrivial effects on dynamics of the spin chain such as ESR.
In the presence of the uniform DM interaction, increase of the magnetic field rotates chirally the spin chain.
This chiral rotation affects selection rules of the electron spin resonance.
It will be interesting to test experimentally the coexistence of the N\'eel and the dimer orders in spin-chain compounds with the fourfold screw symmetry.

\section*{Acknowledgments}

I thank Akira Furusaki, Yusuke Horinouchi, and Tsutomu Momoi for useful discussions.

\appendix

\section{Derivation of effective Hamiltonian}\label{app:H_eff}

This section is devoted to derivation of the low-energy effective Hamiltonian discussed in Sec.~\ref{sec:H_eff} as generically as possible.

\subsection{Framework}\label{app:H_eff_gen}

I consider a Hamiltonian $\mathcal H_0$ whose eigenstates are exactly known.
\begin{align}
    \mathcal H_0 \ket{\phi_n} &= E_n \ket{\phi_n}, \quad (n=0, 1,2, \cdots).
\end{align}
I can assume $E_n \le E_m$ for $n\le m$ without loss of generality.
Adding a perturbation $\lambda V$, I modify the Hamiltonian to
\begin{align}
    \mathcal H
    &= \mathcal H_0 + \lambda V,
    \label{H}
\end{align}
where $\lambda$ is a small parameter that controls the perturbation expansion.
At low energies, effects of the perturbation can be taken into account as a form of the effective Hamiltonian $\mathcal H_{\rm eff}$.
The effective Hamiltonian can be easily obtained up to the second order.

To derive the low-energy effective Hamiltonian $\mathcal H_{\rm eff}$, I focus on the $N$ low-energy eigenstates $\ket{\phi_n}$ with $n=n_0,n_0+1,n_0+2, \cdots, n_0+N-1$ for $n_0 \ge 0$.
One can take $n_0=0$ or $n_0>0$.
The latter case is useful for later application to the free spinless fermion chain.
When applying to the free spinless fermion chain, I assume $|E_n - \epsilon_F| < W$ for $n=n_0, n_0+1, \cdots, n_0+N$, where $\epsilon_F$ is
the Fermi energy and $W>0$ is an energy cutoff.

Each eigenstate defines a projection operator $P_n = \ket{\phi_n}\bra{\phi_n}$ into that eigenstate.
$P_n$ satisfies $P_n \mathcal H_0 = \mathcal H_0 P_n = E_n P_n$.
An operator $P$,
\begin{align}
    P &= \sum_{n=n_0}^{n_0+N-1} P_n,
    \label{P_def}
\end{align}
then projects an arbitrary state into the subspace spanned by the $N$ eigenstates.
$Q=1-P$ projects any state into the supplementary space.

The key idea is to perform a canonical transformation of the Schrieffer-Wolff type on the Hamiltonian~\cite{schriefferwolff, bravyi_sw_transf},
\begin{align}
    \mathcal H' &= e^{\eta}\mathcal H e^{-\eta},
    \label{H'}
\end{align}
where $\eta$ is anti-Hermitian so that $e^{\eta}$ is unitary.
The Schrieffer-Wolff formulation is useful in quantum spin
systems~\cite{kim_deg}.
Here, I briefly review the derivation of the effective Hamiltonian based on the Schrieffer-Wolff formulation to make the paper self-contained.

Two Hamiltonians $\mathcal H$ and $\mathcal H'$ have one-to-one corresponding lists of eigenstates with exactly the same eigenenergies.
An appropriate choice of $\eta$ simplifies the transformed Hamiltonian $\mathcal H'$.
I expand $e^{\eta}$ and determine $\eta$.
\begin{align}
    \mathcal H' 
    &=  \mathcal H + [\eta, \mathcal H] + \frac{1}{2} [\eta, [\eta, \mathcal H]] + \cdots
    \notag \\
    &= \mathcal H_0 + \lambda \Bigl( [\eta_1, \mathcal H_0] + V \biggr) 
    \notag \\
    &\quad + \lambda^2 \biggl( [\eta_2, \mathcal H_0] + [\eta_1, V] + \frac 12 [\eta_1, [\eta_1,\mathcal H_0]] \biggr) + \cdots.
    \label{H'_expand}
\end{align}
In the last line, I expanded $\eta$ around $\lambda=0$:
\begin{align}
    \eta &= \sum_{p=0}^\infty \frac{\lambda^p}{p!}\eta_p.
    \label{eta_expand}
\end{align}
$\eta_p$ is determined so that~\cite{kim_deg}
\begin{align}
    [P, \mathcal H'] &=0.
    \label{cond_H'}
\end{align}
I solve Eq.~\eqref{cond_H'} at each order of $\lambda$.
At the first order, Eq.~\eqref{cond_H'} leads to
\begin{align}
    [P, [\eta_1, \mathcal H_0]] = - [P, V].
    \label{cond_1st}
\end{align}
The anti-Hermitian $\eta_1$ that satisfies Eq.~\eqref{cond_1st} is given by~\cite{kim_deg}
\begin{align}
    \eta_1
    &= \sum_{n=n_0}^{n_0+N-1} \biggl(P_n V \frac{1}{E_n-\mathcal H_0}Q - Q \frac{1}{E_n-\mathcal H_0}V P_n \biggr).
    \label{eta1}
\end{align}
The second order of Eq.~\eqref{cond_H'},
\begin{align}
    [\eta_2, \mathcal H_0] = - X_2,
    \label{cond_2nd}
\end{align}
with $X_2$ being
\begin{align}
    X_2 &= [\eta_1, V] + \frac 12 [\eta_1, [\eta_1, V]],
    \label{X_2}
\end{align}
is similar to the first-order equation \eqref{cond_1st}.
The solution is immediately obtained.
\begin{align}
    \eta_2
    &= \sum_{n=n_0}^{n_0+N-1} \biggl( P_n X_2 \frac{1}{E_n -\mathcal H_0} Q - Q \frac{1}{E_n-\mathcal H_0} X_2 P_n \biggr).
    \label{eta2}
\end{align}

I am now ready to write down the low-energy effective Hamiltonian,
\begin{align}
    \mathcal H_{\rm eff} &= P \mathcal H' P = \sum_{n=0}^\infty \lambda^n \mathcal H_n^{\rm eff},
\end{align}
up to the second order of $\lambda$.
First three terms $\mathcal H_n^{\rm eff}$ for $n=0,1,2$ are shown below.
\begin{align}
    \mathcal H_0^{\rm eff}
    &= P \mathcal H_0 P,
    \label{H_eff_0} \\
    \mathcal H_1^{\rm eff}
    &= P \Bigl( [\eta_1, \mathcal H_0] + V \Bigr) P
    \notag \\
    &= P V P,
    \label{H_eff_1}
\end{align}
where $P [\eta_n, \mathcal H_0]P =0$ holds true for $n=1,2$.
The second-order term $\mathcal H_2^{\rm eff}$ is given by
\begin{align}
    \mathcal H_2^{\rm eff}
    &= P \Bigl( [\eta_2, \mathcal H_0] + X_2 \Bigr) P
    \notag \\
    &= P X_2 P
    \notag \\
    &= P \biggl( [\eta_1, V] + \frac 12 [\eta_1, [\eta_1, \mathcal H_0]] \biggr) P.
\end{align}
One can simplify the last line:
\begin{align}
    \mathcal H_2^{\rm eff}
    &= \frac 12\sum_{n,m} P_n \biggl( V \frac{1}{E_n - \mathcal H_0} QV + V \frac{1}{E_m-\mathcal H_0}Q V\biggr)P_m.
    \label{H_eff_2}
\end{align}
This leads to the following effective Hamiltonian of the second order of $\lambda$:
\begin{align}
    \mathcal H_{\rm eff}
    &= P (\mathcal H_0 +\lambda V) P 
     + \frac{\lambda^2}2 \sum_{n,m =0}^{n_0+N-1}  P_n \biggl( V \frac{1}{E_n - \mathcal H_0} QV 
    \notag \\
    &\quad + V \frac{1}{E_m-\mathcal H_0}Q V\biggr)P_m.
    \label{H_eff}
\end{align}

\subsection{Application to spinless fermion chains}\label{app:H_eff_fermion}

Here I apply the generic formalism of the effective Hamiltonian to the spinless fermion chain \eqref{H_XY_k}.
The low-energy region is defined as Eq.~\eqref{R}.
The operator $P$ is redefined as a projection operator onto the subspace \eqref{R} of the reciprocal space [Eq.~\eqref{R}].
$P$ acts on $c_k$ as follows.
$Pc_kP = c_k$ for $k\in R$ and $Pc _k P=0$ otherwise.
$P$ acts on $c_k^\dag$ in the same manner.
$Q=1-P$ acts on $c_k$ and $c_k^\dag$ as $P c_k Q = Q c_k P = P c_k^\dag Q = Q c_k^\dag P = 0$.
In applying the generic Schrieffer-Wolff formulation to the XY chain \eqref{H_XY_k}, I regard $\mathcal H_0$ and $V$ of Eq.~\eqref{H} as
\begin{align}
    \mathcal H_0
    &= \sum_k \epsilon(k) c_k^\dag c_k, \\
    \lambda V
    &= -\frac{h_4}{\sqrt 2} \sum_k \bigl(
    e^{-\pi i/4} c_k^\dag c_{k+\frac \pi 2} + e^{\pi i/4} c_{k+\frac \pi 2}^\dag c_k \bigr).
\end{align}
The effective Hamiltonian \eqref{H_eff} is then given by
\begin{align}
    \mathcal H_{\rm eff}
    &= \sum_{k\in R} \epsilon(k) c_k^\dag c_k + V',
\end{align}
where $V'=h_4^2 \mathcal H_2^{\rm eff}$ is the second-order term.
Note that the first-order term $\mathcal H_1^{\rm eff}$ vanishes trivially because $P c_{k+\frac \pi 2}^\dag c_k P = 0$ for any $k \in R$ or $k\not\in R$ thanks to the assumption $\Lambda \ll 1$.
The second-order correction $V'$ is calculated as follows.
\begin{widetext}
\begin{align}
    V' &= \frac{h_4^2}{4}\sum_{k,k'} \biggl[
    e^{-\pi i/2} P c_{k+\frac \pi 2}^\dag c_k Q c_{k'+\frac \pi 2}^\dag c_{k'} P \biggl( \frac{1}{\epsilon(k') - \epsilon(k'+\frac{\pi}{2})} + \frac{1}{\epsilon(k+\frac \pi 2) - \epsilon(k)} \biggr)
    \notag \\
    &\qquad + P c_{k+\frac \pi 2}^\dag c_k Q c_{k'}^\dag c_{k'+\frac \pi 2} P \biggl( \frac{1}{\epsilon(k'+\frac \pi 2) -\epsilon(k')} + \frac{1}{\epsilon(k+\frac \pi 2) - \epsilon(k)} \biggr)
    \notag \\
    &\qquad + P c_k^\dag c_{k+\frac \pi 2} Q c_{k'+\frac \pi 2}^\dag c_{k'} P \biggl( \frac{1}{\epsilon(k') - \epsilon(k'+\frac \pi 2)} + \frac{1}{\epsilon(k) - \epsilon(k+\frac \pi 2)} \biggr)
    \notag \\
    &\qquad +  e^{\pi i/2} P c_k^\dag c_{k+\frac \pi 2} Q c_{k'}^\dag c_{k'+\frac \pi 2} P\biggl( \frac{1}{\epsilon(k'+\frac \pi 2) - \epsilon(k')} + \frac{1}{\epsilon(k) - \epsilon(k+\frac \pi 2)}\biggr)
    \biggr]
\end{align}.
One can simplify these projections.
Since the Fermi surface is located at  $k = \pm \pi/2 \mod \pi$, the projection $P c_{k'+\pi}^\dag c_{k'} P$ gives back $c_{k'+\pi}^\dag c_{k'}$ itself for $k'\in R$ and zero otherwise.
In the end, I obtain
\begin{align}
    V'
    &= -i\frac{h_4^2}{4}\sum_{k\in R} (c_{k+\pi}^\dag c_k - c_{k}^\dag c_{k+\pi}) \biggl( \frac{1}{\epsilon(k) - \epsilon(k+\frac{\pi}{2})} + \frac{1}{\epsilon(k+\pi) - \epsilon(k+\frac \pi 2)} \biggr)
    \notag \\
    &\qquad + \frac{h_4^2}{2} \sum_{k\in R} c_k^\dag c_k  \biggl( \frac{1}{\epsilon(k) - \epsilon(k-\frac \pi 2)} + \frac{1}{\epsilon(k) -\epsilon(k+\frac \pi 2)} \biggr).
    \label{V'}
\end{align}
\end{widetext}
The first line of Eq.~\eqref{V'} is the the bond alternation for $k\approx \pm k_F$~\cite{giamarchi_book} and the second line is a small correction to the Zeeman energy.

\section{Electron spin resonance}\label{app:esr}

Here, I describe how the uniform DM interaction affects the ESR spectrum.
In this Appendix, I start with the spin chain model \eqref{H_exp_rot} with $\lambda=1$.
Namely, I consider the spin chain with the following Hamiltonian:
\begin{align}
    \tilde{\mathcal H}_{\rm exp}
    &= J \sum_j \bm S_j \cdot \bm S_{j+1} + h_0 \sum_j S_j^z 
    \notag \\
    &\qquad + h_2 \sum_j (-1)^j S_j^x +\frac{2J\delta_\perp}{3}\sum_j (-1)^j \bm S_j \cdot \bm S_{j+1}
    \notag \\
    &\qquad + D_u \sum_j (S_j^z S_{j+1}^x - S_j^x S_{j+1}^z).
    \label{H_app_exp}
\end{align}
The ESR spectrum is obtained from the $q=0$ part of the dynamical correlation function $\braket{S^aS^b}(q,\omega)$ for $a,b=x,y,z$.
I can obtain selection rules of the ESR spectrum by relating the $q=0$ part of the spin, $S^a_{q=0}$, where
\begin{align}
    S^a_{q} :=  \sum_j e^{iqj} S_j^a,
\end{align}
to the boson fields of the effective field theory.

Let us bosonize the spin chain by using the non-Abelian bosonization formula~\cite{gogolin_book, affleck_nonabelian},
\begin{align}
    \bm S_j &= \bm J_R + \bm J_L -\frac{ib_0}2 (-1)^j \tr(g\bm \sigma),
\end{align}
where $\bm J_R$, $\bm J_L$, and $g$ are defined as
\begin{align}
    J_R^z &= - \frac{i}{\sqrt{2\pi}} \bar{\partial} \varphi_R, \\
    J_L^z &= \frac{i}{\sqrt{2\pi}} \partial \varphi_L, \\
    J_R^\pm &= \frac{1}{2\pi} e^{\pm i \sqrt{8\pi}\varphi_R}, \\
    J_L^\pm &= \frac{1}{2\pi} e^{\mp i \sqrt{8\pi}\varphi_L}, \\
    g &=
    \begin{pmatrix}
    e^{i\sqrt{2\pi}\phi} & ie^{-i\sqrt{2\pi}\theta} \\
    ie^{i\sqrt{2\pi}\theta} & e^{-i\sqrt{2\pi} \phi}
    \end{pmatrix}.
\end{align}
Here $\varphi$ and $\varphi$ at a position $x$ and a time $t$ are related to $\varphi_R$ and $\varphi_L$ through
\begin{align}
    \phi(x,t) &= \varphi_R(x-vt) + \varphi_L(x+vt), \\
    \theta(x,t) &= \varphi_R(x-vt) - \varphi_L(x+vt).
\end{align}
The derivatives $\partial$ and $\bar{\partial}$ are abbreviations of the following derivatives.
\begin{align}
    \partial &= \frac{-i}{2}(\partial_x + v^{-1}\partial_t), \\
    \bar{\partial} &= \frac{i}{2}(\partial_x - v^{-1}\partial_t).
\end{align}
Boson fields $\phi$ and $\theta$ are subject to equal-time commutation relations,
\begin{align}
    [\phi(x), \theta(y)] = iY(y-x),
\end{align}
with a step function,
\begin{align}
    Y(y-x) &=
    \left\{
    \begin{array}{ccc}
         1, & & (y>x),  \\
         0, & & (y<x), \\
         1/2, & & (y=x). 
    \end{array}
    \right.
\end{align}
$S_j^a$ for $a=x,y,z$ are thus bosonized as~\cite{mudry_xyz}
\begin{align}
    S_j^x &= b_0 \cos(\sqrt{2\pi}\theta) + ib_1\sin(\sqrt{2\pi}\theta) \sin(\sqrt{2\pi}\phi), \\
    S_j^y &= b_0 \sin(\sqrt{2\pi}\theta) + ib_1 \cos(\sqrt{2\pi}\theta) \sin(\sqrt{2\pi}\phi),
    \\
    S_j^z &= \frac{1}{\sqrt{2\pi}} \partial_x \phi + b_0 \sin(\sqrt{2\pi}\phi).
\end{align}
In the non-Abelian bosonization laugage, the Hamiltonian \eqref{H_app_exp} is expressed as
\begin{align}
    \tilde{\mathcal H}_{\rm exp}
    &= \frac{2\pi v}{3} \int dx\, (\bm J_R \cdot \bm J_R+\bm J_L \cdot \bm J_L)
    \notag \\
    &\qquad + h_0 \int dx \, (J_R^z + J_L^z) + \gamma D_u \int dx (J_R^y - J_L^y)
    \notag \\
    &\qquad - \frac{ib_0h_2}{2} \int dx \, \tr(g\sigma^x) + \frac{d_{xy}J\delta_\perp}{3} \int dx \, \tr(g).
\end{align}
In order to combine terms on the second line, I perform the chiral rotation \eqref{rot}.
The chiral rotation turns the Hamiltonian into
\begin{align}
    \tilde{\mathcal H}_{\rm exp}
    &= \frac{2\pi v}{3} \int dx \, (\bm M_R \cdot \bm M_R + \bm M_L \cdot \bm M_L)
    \notag \\
    &\qquad + t_\phi \int dx \, (M_R^z + M_L^z)
    \notag \\
    &\qquad -\frac{ib_0 h_2}{2} \int dx \,[\tr(g'\sigma^x)\cos \theta_R +i \tr(g') \sin \theta_R]
    \notag \\
    &\qquad +\frac{d_{xy}J\delta_\perp}{3} \int dx \, [\tr(g') \cos \theta_R + i \tr(g'\sigma^x) \sin \theta_R].
    \label{H_app_exp_crot}
\end{align}
$\bm M_R$, $\bm M_L$, and $g'$ are related to U(1) bosons $\Phi = \varphi'_R + \varphi'_L$ and $\Theta = \varphi'_R - \varphi'_L$ in analogy with $\bm J_R$, $\bm J_L$, and $g$.
I can eliminate the Zeeman energy $t_\phi (M_R^z + M_R^z)$ by shifting
\begin{align}
     \varphi'_R &\to \varphi'_R -\frac{t_\phi}{v\sqrt{8\pi}}x, \\
    \varphi'_L &\to \varphi'_L - \frac{t_\phi}{v\sqrt{8\pi}}x.
\end{align}
This shift affect $\bm M_R$, $\bm M_L$, and $g'$ as follows.
\begin{align}
    M_R^z &= - \frac{t_\phi}{4\pi v} - \frac{i}{\sqrt{2\pi}}\bar\partial \varphi'_R, \\
    M_L^z &= -\frac{t_\phi}{4\pi v} + \frac{i}{\sqrt{2\pi}}\partial \varphi'_L, \\
    M_R^\pm &= e^{\mp i t_\phi x/v} e^{\pm i \sqrt{8\pi}\varphi'_R}, \\
    M_L^\pm &= e^{\pm i t_\phi x/v} e^{\mp i \sqrt{8\pi}\varphi'_L}, \\
    g'&=
    \begin{pmatrix}
        e^{-it_\phi x/v} e^{i\sqrt{2\pi}\Phi} & ie^{-i\sqrt{2\pi}\Theta} \\
        ie^{i\sqrt{2\pi}\Theta} & e^{it_\phi x/v} e^{-i\sqrt{2\pi}\Phi}
    \end{pmatrix}.
\end{align}
The shift introduces incommensurate oscillations to the Hamiltonian \eqref{H_app_exp_crot}.
Here, I assume an inequality,
\begin{align}
    \frac{v}{\Delta} \ll \frac{v}{t_\phi}
\end{align}
This condition guarantees that the incommensurate oscillation is negligible (cf. Secs.~\ref{sec:rg} and \ref{sec:nonAbelian}).
At low fields $h_0/J \ll 1$, this inequality reads
\begin{align}
    (h_0/J)^{2/3} \gg t_\phi/J.
    \label{ineq_tphi}
\end{align}
Under this condition \eqref{ineq_tphi}, I can safely discard the incommensurate oscillation with the wave number $t_\phi/v$ in the Hamiltonian.
Note that $t_\phi$ must be kept in the relations between operators and quantum fields.
The Hamiltonian is thus given by
\begin{align}
    \tilde{\mathcal H}_{\rm exp}
    &\approx \frac{2\pi v}{3} \int dx \, (\bm M_R \cdot \bm M_R + \bm M_L \cdot \bm M_L)
    \notag \\
    &\qquad -\frac{ib_0 h_2}{2} \int dx \,[\tr(g'\sigma^x)\cos \theta_R +i \tr(g') \sin \theta_R]
    \notag \\
    &\qquad +\frac{d_{xy}J\delta_\perp}{3} \int dx \, [\tr(g') \cos \theta_R + i \tr(g'\sigma^x) \sin \theta_R].
\end{align}
Here, as I did in Sec.~\ref{sec:nonAbelian}, I rotate the system by $\pi$ around the $y$ axis:
$(\sigma^x, \, \sigma^y, \, \sigma^z) \to (\sigma^z, \, \sigma^y,\, -\sigma^x)$.
The $\pi$-rotated Hamiltonian finally becomes simple.
\begin{align}
    \tilde{\mathcal H}_{\rm exp}
    &=\frac{v}{2} \int dx \, [v^{-2}(\partial_t \phi)^2 + (\partial_x \phi)^2 ]
    \notag \\
    &\qquad +G \int dx \, \cos(\sqrt{2\pi} \Phi +\theta_R -\alpha'),
    \label{H_app_exp_SG}
\end{align}
with
\begin{align}
    G &= \sqrt{(b_0 h_2)^2 + (2d_{xy} J \delta_\perp /3)^2},
    \label{G} \\
    \alpha' &= \tan^{-1}\biggl( \frac{3b_0h_2}{2d_{xy} J\delta_\perp}\biggr).
    \label{alpha'}
\end{align}

Let us relate the spin $\bm S_j$ in the original coordinate frame to the $\Phi$ and $\Theta$ fields in Eq.~\eqref{H_app_exp_SG}, recalling all the chiral and nonchiral rotations performed.
\begin{widetext}
\begin{align}
    S_j^x
    &= M_R^x + M_L^x -\frac{ib_0}{2}[\tr(g'\sigma^x) \cos \theta_R +i \tr (g') \sin \theta_R]
    \notag \\
    &= \frac{1}{\sqrt{2\pi}}( \partial_x \Phi ) \cos(t_\phi x/v) +ib_1 \cos(\sqrt{2\pi}\Theta) \sin(\sqrt{2\pi}\Phi) \sin (t_\phi x/v) + b_0 (-1)^j \sin(\sqrt{2\pi}\Phi + \theta_R),
    \label{Sx_app}
    \\
    S_j^y
    &= (M_R^y + M_L^y) \cos \theta_R -(M_R^z - M_L^z)\sin \theta_R -\frac{ib_0}{2}(-1)^j \tr(g'\sigma^y)
    \notag \\
    &= \biggl(ib_1 \cos(\sqrt{2\pi}\Theta) \sin(\sqrt{2\pi}\Phi) \cos(t_\phi x/v) -\frac{1}{\sqrt{2\pi}}(\partial_x \Theta) \sin(t_\phi x/v)\biggr)\cos \theta_R 
    \notag \\
    &\qquad -ib_1 \cos(\sqrt{2\pi}\Theta) \cos(\sqrt{2\pi}\Phi) \sin \theta_R + b_0 (-1)^j \sin(\sqrt{2\pi}\Theta), 
    \label{Sy_app}
    \\
    S_j^z
    &= (M_R^y - M_L^y) \sin \theta_R + (M_R^z + M_L^z) \cos \theta_R -\frac{ib_0}{2} (-1)^j \tr(g'\sigma^z)
    \notag \\
    &=-\frac{t_\phi}{2\pi v} \cos\theta_R +\biggl(ib_1 \sin(\sqrt{2\pi}\Theta) \cos(\sqrt{2\pi}\Phi) \cos( t_\phi x/v) +\frac{1}{\sqrt{2\pi}}(\partial_x \Phi) \cos(t_\phi x/v) \biggr)\sin \theta_R
    \notag \\
    &\qquad 
    -ib_1 \sin(\sqrt{2\pi}\Theta) \sin(\sqrt{2\pi}\Phi)  +b_0 (-1)^j \sin(\sqrt{2\pi}\Phi - t_\phi x/v).
    \label{Sz_app}
\end{align}
Note that the $\pi$ rotation was performed on the rightmost hand sides of Eqs.~\eqref{Sx_app}, \eqref{Sy_app}, and \eqref{Sz_app}.
The transverse dimer order is expressed as
\begin{align}
    (-1)^j (S_j^x S_{j+1}^x + S_j^y S_{j+1}^z)
    &= \frac{d_{xy}}{2} (\tr(g') \cos \theta_R + i\tr(g\sigma^x) \sin \theta_R)
    \notag \\
    &= d_{xy} \cos(\sqrt{2\pi}\Phi + \theta_R).
\end{align}
\end{widetext}

I can confirm that the spin chain model \eqref{H_app_exp} has the N\'eel and dimer orders.
\begin{align}
    N_x
    &\propto (G/J)^{2/3} \sin\alpha', \\
    D_\perp
    &\propto (G/J)^{2/3} \cos \alpha',
\end{align}
in analogy with Eqs.~\eqref{Nx_SG} and \eqref{D_perp_SG}.

A list of low-energy excitations created by operators on the rightmost hand sides of Eqs~\eqref{Sx_app}, \eqref{Sy_app}, and \eqref{Sz_app} is available~\cite{lukyanov_formfactor1, lukyanov_formfactor2, babujian_formfactor}.
Vertex operator operators $e^{iqx}e^{\pm i\sqrt{2\pi}\Theta}$ create solitons and antisolitons with an excitation energy,
\begin{align}
    E_s(q) 
    &=\sqrt{(vq)^2 + {\Delta_s}^2}, \\
    \Delta_s
    &= \frac{2v}{\sqrt{\pi}}\frac{\Gamma(1/6)}{\Gamma(2/3)} \biggl( \frac{\pi }{2v}\frac{\Gamma(3/4)}{\Gamma(1/4)} G \biggr)^{2/3}.
    \label{Delta_s}
\end{align}
Other vertex operators $e^{iqx} e^{\pm i \sqrt{2\pi}\Phi}$ create breathers, bound states of a soliton and an antisoliton, with an excitation energy,
\begin{align}
    E_n(q) &= \sqrt{(vq)^2 + \Delta_n^2}, \\
    \Delta_n &= 2\Delta_s \sin\biggl(\frac{n\pi \xi}{2}\biggr).
\end{align}
Here, $\xi = 1/(8K-1) =3$.
The index $n$ takes values of $n=1,2,3$.

I can now predict the ESR frequency caused by $S_{q=0}^a$ for $a=x,y,z$.
For example, $\cos(\sqrt{2\pi}\Theta) \cos(\sqrt{2\pi}\Phi)$ in  $S_{q=0}^y$ yields the resonance peak at 
\begin{align}
    \omega =\Delta_s \propto G^{2/3}.
    \label{res1}
\end{align}
Though in the $h_0/J \to 0$ limit, this resonance frequency follows a simple power law $\omega \propto (h_0/J)^{2/3}$, it will be a complicated function of $h_0$ in general.
Another interesting term is $\sin(\sqrt{2\pi}\Theta) \cos(\sqrt{2\pi}\Phi) \cos (t_\phi x/v)$ in $S_{q=0}^z$.
This term yields resonance peaks at
\begin{align}
    \omega &= \sqrt{{t_\phi}^2 + M^2},
\end{align}
where $M$ can be $\Delta_s$ or $\Delta_n$ for $n=1,2,3$.

The selection rule is also affected by the canonical transformation \eqref{H'}.
Precisely speaking, the left hand sides of Eqs.~\eqref{Sx_app}, \eqref{Sy_app}, and \eqref{Sz_app} should be denoted as $\tilde S_j^a$ for $a=x,y,z$.
Here, $\tilde S_j^a$ is defined as
\begin{align}
    \tilde S_j^a &= e^{\eta} S_j^a e^{-\eta}
    \notag \\
    &= S_j^a + [\eta_1, S_j^a] + [\eta_2, S_j^a] + \frac 12 [\eta_1, [\eta_1, S_j^a]] + \cdots.
\end{align}
$\eta_1$ and $\eta_2$ create particle-hole excitations with $q=\pm \pi/2$ and $\pi$, respectively, when applied to the TL-liquid ground state.
Such a mixing of different wave numbers will allow ESR to detect $q=\pm \pi/2$ and $q=\pi$ excitations.
Excitations with $q=\pi$ can be read from the staggered terms of Eqs.~\eqref{Sx_app}, \eqref{Sy_app}, and \eqref{Sz_app}, which are similar to those with $q=0$.

\bibliography{ref.bib}

\end{document}